\documentclass[%
 reprint,
 amsmath,amssymb,
 aps,
 pra,
]{revtex4-2}

\usepackage{braket}
\usepackage{graphicx}
\usepackage{mathtools}
\usepackage{booktabs}
\usepackage[colorlinks=true, allcolors=blue]{hyperref}
\newcommand{\xx}{\mathbf{X}}
\newcommand{\DD}{\mathcal{D}}
\newcommand{\EE}{\mathcal{E}}
\newcommand{\bs}[1]{\boldsymbol{#1}}
\newcommand{\yy}{\mathbf{Y}}
\newcommand{\nc}{\begin{pmatrix}N\\2\end{pmatrix}}
\DeclareMathOperator{\Tr}{Tr}
\newcommand{\conf}[1]{\{#1\}}

\begin{document}

\title{Exact solutions to the quantum many-body problem using the geminal density matrix} 

\author{Nicholas Cox*}
\affiliation{ICFO--Institut de Ciencies Fotoniques, The Barcelona Institute of Science and Technology, 08860 Castelldefels, Barcelona, Spain}
\affiliation{The College of Optics and Photonics (CREOL), University of Central Florida, Orlando, Florida 32816, USA}
\email{Nickcox@knights.ucf.edu}
\date{\today}

\begin{abstract}
It is virtually impossible to directly solve the Schr\"odinger equation for a many-electron wave function due to the exponential growth in degrees of freedom with increasing particle number. The two-body reduced density matrix (2-RDM) formalism reduces this coordinate dependence to that of four particles irrespective of the wave function's dimensionality, providing a promising path to solve the many-body problem. Unfortunately, errors arise in this approach because the 2-RDM cannot practically be constrained to guarantee that it corresponds to a valid wave function. Here we approach this so-called $N$-representability problem by expanding the 2-RDM in a complete basis of two-electron wave functions and studying the matrix formed by the expansion coefficients. This quantity, which we call the geminal density matrix (GDM), is found to evolve in time by a unitary transformation that preserves $N$-representability. This evolution law enables us to calculate eigenstates of strongly correlated systems by a fictitious adiabatic evolution in which the electron-electron interaction is slowly switched on. We show how this technique is used to diagonalize atomic Hamiltonians, finding that the problem reduces to the solution of $\sim N(N-1)/2$ two-electron eigenstates of the Helium atom on a grid of electron-electron interaction scaling factors.
\end{abstract}

\maketitle

\section{Introduction}\label{sec:Intro}
In 1955, L\"owdin  \cite{lowdinQuantumTheoryManyparticle1955a,lowdinQuantumTheoryManyParticle1955,lowdinQuantumTheoryManyparticle1955b} and Mayer \cite{mayerElectronCorrelation1955} presented similar methods to express the ground state energy of a many-electron quantum system as a functional of the two-body reduced density matrix ($2$-RDM). Their work inspired belief in the feasibility of solving complex many-body problems with an effective two-particle analysis \cite{coulsonPresentStateMolecular1960}. However, early calculations significantly underestimated experimental ground state energies because the $2$-RDM was not adequately constrained to ensure that it represents a valid many-body wave function. Full determination of these constraints, known as the $N$-representability conditions \cite{colemanStructureFermionDensity1963}, would indeed yield a method to reduce the many-body problem to an effective two-particle system. The development of new constraints continues to improve the accuracy of 2-RDM calculations, but the general problem remains unsolved. 

As detailed in Ref. \cite{mazziottiTwoElectronReducedDensity2012}, modern $2$-RDM analysis primarily employs one of two methods. The first begins by deriving a contracted Schr\"odinger equation (CSE) that computes the energy as a function of the two and four-body reduced density matrices ($4$-RDM) \cite{nakatsujiEquationDirectDetermination1976,nakatsujiDirectDeterminationQuantumMechanical1996,mazziottiContractedSchrodingerEquation1998,mazziottiComparisonContractedSchrodinger1999}. Approximating the $4$-RDM as a function of the $2$-RDM allows one to solve the CSE for a set of candidate eigenstates from which the physically valid states are selected by imposing $N$-representability conditions. The second technique, density matrix variational theory (DMVT), aims to directly minimize the energy as a functional of the $2$-RDM \cite{garrodReductionParticleVariational1964,mihailovicVariationalApproachDensity1975,mazziottiUncertaintyRelationsReduced2001,nakataVariationalCalculationsFermion2001,mazziottiRealizationQuantumChemistry2004}. The most successful applications of DMVT use a convex optimization scheme called semi-definite programming \cite{vandenbergheSemidefiniteProgramming1996,nakataVariationalCalculationsFermion2001} in which the $N$-representability conditions are included by a set of positivity conditions that restrict the search space for solutions.

Here we take a conceptually simpler approach that begins by expanding the $2$-RDM in a basis of two-electron \textit{geminal} \cite{surjanIntroductionTheoryGeminals1999} eigenstates. The resulting expansion coefficients are collected into a quantity we call the geminal density matrix (GDM) that can be used to compute many-body observables from effective two-body operators. The technique was first introduced by Bopp in an attempt to calculate the ground state energy of selected ions \cite{boppAbleitungBindungsenergieNTeilchenSystemen1959}. Although his method was exact, his results differed quite significantly from experimental ground state energies. These errors were later attributed to the non-$N$-representability of the assumed ground state matrix \cite{colemanStructureFermionDensity1963}. Until now, very little work has been done to advance this matrix-based approach.

We will detail how the GDM formalism enables us to calculate the stationary states of a general many-body Hamiltonian with two-body (electron-electron) interactions. To do so, we need to place Bopp's work on a solid theoretical foundation to make sense of $N$-representability in the context of GDMs. Most importantly, we find that the GDM must evolve unitarily in time by the Liouville-Von Neumann equation in order to produce the same expectation values as the time-dependent $N$-electron wave function. Since the wave function is only useful insofar as it generates observables, any matrix that reproduces these quantities is clearly a faithful representation of the quantum state.

Because the equation of motion preserves $N$-representability, we find it useful to examine a Hamiltonian that slowly switches on the electron-electron interaction by some time-dependent scaling. We show by the adiabatic theorem that eigenstates of the non-interacting (initial) system evolve into those of the interacting (final) system. In this way, we find it possible to construct $N$-electron stationary states using $\sim N(N-1)/2$ eigenstates of an effective two-electron Hamiltonian computed on a grid of interaction scaling strengths. 

As an example, we show that the effective Hamiltonian for an arbitrary atom or ion reduces to a coordinate-scaled Helium Hamiltonian with some specific electron-electron interaction strength. The result is that all atomic electron energy eigenstates can be found strictly from the solution to this Helium atom problem. While atoms provide the simplest use case, the formalism presented here applies equally well to molecular and solid state systems.

The paper is organized as follows. Section \ref{sec:Density} introduces the $2$-RDM then expands it in a two-electron basis to define the GDM. Section \ref{sec:States} gives examples of valid density matrices that serve as the starting point for eigenstate calculations. Section \ref{sec:Time} derives the unitary time evolution law of the GDM, which Section \ref{sec:Evolution} applies to the cases in which the electron-electron interaction is switched on suddenly or slowly. The latter leads to the adiabatic theorem, which is exploited in section \ref{sec:Solve} to solve the many-body Schr\"odinger equation.

Appendix \ref{sec:Properties} derives the four necessary $N$-representability constraints listed in Section \ref{sec:Density}. Appendix \ref{sec:Basis} defines the matrix transformation imposed by a change of geminal basis, and Appendix \ref{sec:Alternate} provides an alternate derivation of the GDM equation of motion found in Section \ref{sec:Time}.

Hartree atomic units with $\hbar = e = m_0 = 1/(4\pi \epsilon_0) = 1$ will be used throughout.

\section{The geminal density matrix}\label{sec:Density}
This section presents the GDM as the basic quantity needed to calculate the observables of a many-body quantum system. It begins with an introductory derivation of the $2$-RDM and proceeds to define the GDM by expanding the $2$-RDM in a two-electron basis. The section ends by presenting a short list of necessary $N$-representability conditions that are derived in Appendix \ref{sec:Properties}.

\subsection{The 2-RDM}

The state of a given $N$-electron system is fully described by its wave function $\Psi(\mathbf{x}_1,\dots,\mathbf{x}_N)$, where we combined spatial and spin degrees of freedom into the symbol
\begin{equation}
\mathbf{x}_i = \mathbf{r}_i\sigma_i.
\end{equation}
The wave function is the position representation $\Psi(\mathbf{x}_1,\dots,\mathbf{x}_N) = \braket{\mathbf{x}_1,\dots,\mathbf{x}_N|\alpha}$ of the abstract vector $\ket{\alpha}$, and it must be anti-symmetric under coordinate exchange $\mathbf{x}_i \leftrightarrow \mathbf{x}_j$ for any pair $i$ and $j$. We impose the normalization condition
\begin{equation}\label{eq:norm}
	\int |\Psi(\mathbf{x}_1,\dots,\mathbf{x}_N)|^2 \prod_{i=1}^N d\mathbf{x}_i = 1,
\end{equation}
with the integral over $d\mathbf{x}$ including a sum over spin coordinates by
\begin{equation}\label{eq:spinsum}
	\int f(\mathbf{x}) d\mathbf{x} = \sum_{\sigma} \int f(\mathbf{r}\mathbf{\sigma}) d\mathbf{r}.
\end{equation}

Observables are represented by $N$-body linear operators with position representation $A(\mathbf{x}_1,\dots,\mathbf{x}_N) = \braket{\mathbf{x}_1,\dots,\mathbf{x}_N|\hat{A}|\mathbf{x}_1,\dots,\mathbf{x}_N}$. An operator must be symmetric under particle exchange so that it maps anti-symmetrized wave functions to anti-symmetrized wave functions. In general, $A$ contains one-body components acting on the coordinates of single particles and two-body components acting on pairs. Denoting one-body contributions by lower case letters and two-body by upper case, $A$ is expressed in the position representation as
\begin{equation}\label{eq:sop}
	A(\mathbf{x}_1,\dots,\mathbf{x}_N) = \sum_k a_1(\mathbf{x}_k) + \sum_{i,j>i}A_2(\mathbf{x}_i,\mathbf{x}_j).
\end{equation}
We will find it convenient to combine Eq. \ref{eq:sop} into the single sum over pairs
\begin{equation}\label{eq:Atot}
	A(\mathbf{x}_1,\dots,\mathbf{x}_N) = \sum_{i,j>i} A(\mathbf{x}_i,\mathbf{x}_j),
\end{equation}
with the summand $A(\mathbf{x}_i,\mathbf{x}_j)$ defined by
\begin{equation}\label{eq:Asep1}
	A(\mathbf{x}_i,\mathbf{x}_j) = A_1(\mathbf{x}_i,\mathbf{x}_j) + A_2(\mathbf{x}_i,\mathbf{x}_j).
\end{equation}
The term $A_1(\mathbf{x}_i,\mathbf{x}_j)$ is the promotion of a one-body operator to act on pairs
\begin{equation}\label{eq:Aeff1}
	A_1(\mathbf{x}_i,\mathbf{x}_j) =  \frac{a_1(\mathbf{x}_i) + a_1(\mathbf{x}_j)}{N-1},
\end{equation}
where the denominator divides out the overcounting that occurs in the pair sum. 

Using Eq. \ref{eq:Atot}, we calculate the expectation value of $A$ by the inner product
\begin{align}\label{eq:AA1}
	\braket{A} &= \sum_{i,j>i}\int \Psi^*(\mathbf{x}_1,\dots,\mathbf{x}_N) \nonumber \\ 
	&\times A(\mathbf{x}_i,\mathbf{x}_j)\Psi(\mathbf{x}_1,\dots,\mathbf{x}_N)\prod_{i=1}^N d\mathbf{x}_i.
\end{align}
Each integral in the pair sum is found to be equivalent by first swapping the names $\mathbf{x}_i \leftrightarrow \mathbf{x}_1$ and $\mathbf{x}_j \leftrightarrow \mathbf{x}_2$ then permuting the same pairs within the argument lists of $\Psi$ and $\Psi^*$. Since each wave function changes sign under coordinate permutation, the argument swapping leaves the integrand unchanged.

Combining like variables by the shorthand $\xx = \mathbf{x}_1,\mathbf{x}_2$ and $\yy = \mathbf{x}_3,\dots,\mathbf{x}_N$, we find that
\begin{equation}\label{eq:A11}
	\braket{A} = \nc\int \Psi^*(\xx,\yy)A(\xx)\Psi(\xx,\yy)d\xx d\yy,
\end{equation}
where $d\xx = d\mathbf{x}_1 d\mathbf{x}_2$ and $d\yy = d\mathbf{x}_3\cdots d\mathbf{x}_N$. The prefactor, equal to $N(N-1)/2$, is the number of equivalent integrals in the pair sum. The operator $A(\xx)$ is exactly the contribution of the single pair in Eq. \ref{eq:Asep1},
\begin{align}\label{eq:A2}
	A(\xx) = \frac{a_1(\mathbf{x}_1) + a_2(\mathbf{x}_2)}{N-1} + A_2(\xx).
\end{align}

Noting that $A(\xx)$ does not depend on $\yy$, we would like to remove it from the integral over $d\yy$. We cannot factor out $A(\xx)$ directly because it may contain derivatives that act on the integral by the chain rule. We circumvent this problem by introducing a set of primed coordinates $\xx'$ upon which the operator $A(\xx')$ is taken to act. This definition permits the re-formulation of Eq. \ref{eq:A11} as
\begin{align}\label{eq:A1}
	\braket{A} = &\nc\int d\xx d\xx' \delta(\xx' - \xx) \nonumber \\
	&\times A(\xx')\int \Psi^*(\xx,\yy)\Psi(\xx',\yy)d\yy.
\end{align}
In Eq. \ref{eq:A1} we defined the delta function
\begin{equation}
	\delta(\xx' - \xx) = \prod_{i=1}^N \delta(\mathbf{r}_i' - \mathbf{r}_i)\delta_{\sigma_i',\sigma_{i}}
\end{equation}
that includes Dirac delta functions for position and Kronecker delta functions for spin polarization. 

Finally, we define the $2$-RDM
\begin{equation}\label{eq:rho1}
	\rho(\xx,\xx') = \nc\int \Psi^*(\xx,\yy)\Psi(\xx',\yy)d\yy,
\end{equation}
so that Eq. \ref{eq:A1} simplifies to
\begin{equation}\label{eq:At1}
	\braket{A} = \int d\xx d\xx' \delta(\xx' - \xx) A(\xx')\rho(\xx,\xx').
\end{equation}
Eqs. \ref{eq:rho1} and \ref{eq:At1} provide an enormous complexity reduction compared to Eq. \ref{eq:AA1}, as they depend only on $\xx$ and $\xx'$ regardless of the number of particles under investigation.

\subsection{From the 2-RDM to the GDM}

We will now represent the $2$-RDM in matrix form, starting by expressing it as the position representation of a non-local two-body linear operator $\hat{D}$:
\begin{equation}\label{eq:rhopos}
	\rho(\xx,\xx') = \braket{\xx'|\hat{D}|\xx}.
\end{equation}
Introducing a complete set of geminal basis states $\ket{i}$ with wave functions $\psi_i(\xx) = \braket{\xx|i}$, we expand $\hat{D}$ by inserting two resolutions of the identity $1 = \sum_{i}\ket{i}\bra{i}$ to find
\begin{equation}\label{eq:Dab}
	\hat{D} = \sum_{mn}\ket{m}\braket{m|\hat{D}|n}\bra{n}.
\end{equation}
Defining $D_{mn} = \braket{m|\hat{D}|n}$ and taking the position representation by pre-multiplying $\bra{\xx'}$ and post-multiplying $\ket{\xx}$, the $2$-RDM as expressed in Eq. \ref{eq:rhopos} takes the form
\begin{equation}\label{eq:rhoexp}
	\rho(\xx,\xx') = \sum_{mn}D_{mn}\psi_n^*(\xx)\psi_m(\xx').
\end{equation}
We can prove the validity of this expansion by construction; applying $\int \psi_{n'}(\xx) \psi_{m'}^*(\xx') d\xx d\xx'$ to either side of Eq. \ref{eq:rhoexp} isolates element $D_{m'n'}$ as a functional of the $2$-RDM. See Eq. \ref{eq:Dmn1} in Appendix \ref{sec:Properties} for more detail.
 
Inserting Eq. \ref{eq:rhoexp} into Eq. \ref{eq:At1} yields
\begin{equation} \label{eq:Ata}
	\braket{A} =  \sum_{mn}D_{mn} A_{nm},
\end{equation}
where
\begin{equation}
	A_{nm} = \int \psi_n^*(\xx)A(\xx)\psi_m(\xx) d\xx.
\end{equation}
We now define the GDM as the matrix $\mathbf{D}$ of coefficients $D_{mn}$. Similarly defining $\mathbf{A}$ to have coefficients $A_{mn}$, Eq. \ref{eq:Ata} reduces to the matrix form
\begin{equation}\label{eq:tr1}
	\braket{A} = \Tr[\mathbf{D}\mathbf{A}].
\end{equation}

Eq. \ref{eq:tr1} was derived without approximation so it exactly reproduces any observable quantity given by the many-body wave function. Thus, knowledge of the matrix $\mathbf{D}$ is equivalent to knowledge of $\Psi(\mathbf{x}_1,\dots,\mathbf{x}_N)$.

Although we can always generate a unique GDM from the many-body wave function, the process can not be reversed to construct the wave function from the GDM. As a result, it is necessary to constrain $\mathbf{D}$ so that it is guaranteed to satisfy $N$-representability. In Appendix \ref{sec:Properties} we derive the following four necessary conditions:
\begin{subequations}
\label{eq:ruless}
\begin{align}
    \mathbf{D} &= \mathbf{D}^\dagger \label{eq:hermitian0}\\
    0\leq &D_{nn} \leq 1 \label{eq:occupation0}\\
    \Tr[\mathbf{D}] &= \begin{pmatrix}
    N \\ 2
    \end{pmatrix}  \label{eq:trace0}\\
     0\leq \Tr[\mathbf{D}^2] &\leq \begin{pmatrix}
    N \\ 2
    \end{pmatrix}. \label{eq:tracesquared0}
\end{align}
\end{subequations}
Condition (\ref{eq:hermitian0}) says the matrix must be Hermitian. We can interpret rule (\ref{eq:occupation0}) as the maximum occupation number for a given geminal and (\ref{eq:trace0}) as the total number of electron pairs present in the wave function. The final expression, (\ref{eq:tracesquared0}), is derivable from the first three and provides a way to distinguish states in a manner that is invariant under unitary matrix transformation. In fact, we will find that all matrices of interest in this work satisfy the strict equality $\Tr[\mathbf{D}^2] = N(N-1)/2$.

In Section \ref{sec:States} we find that Eq. \ref{eq:ruless} is an insufficient set of constraints because it is possible to define a non-$N$-representable GDM that obeys these rules.

\section{Matrix Examples}\label{sec:States}
This section begins with examples of simple GDMs followed by a demonstration of the $N$-representability problem. Subsection \ref{sec:real} explores the application of these matrices to solve for the stationary states of interacting many-body Hamiltonians. It is found that solving such systems may be possible through a time-dependent analysis.

\subsection{Matrices with a non-interacting geminal basis}\label{sec:slater}

Using an orthonormal basis of one-electron wave functions $\phi_i(\mathbf{x})$, we can build $N$-electron Slater determinants by
\begin{equation}\label{eq:basis1}
	\Psi_{\conf{\alpha}}(\mathbf{x}_1,\dots,\mathbf{x}_N) = \hat{S}_- \prod_{i=1}^N \phi_{\alpha_i}(\mathbf{x}_{i}).
\end{equation}
In Eq. \ref{eq:basis1}, we defined a configuration $\conf{\alpha}$ to be an ordered collection of integers $\alpha_i$ that specify the single-particle eigenstates included in a given product. The operator $\hat{S}_-$ transforms the product into an anti-symmetrized wave function by the determinant operator in Eqs. \ref{eq:symop1} and \ref{eq:slater}.

Per Eq. \ref{eq:rhoexp}, the GDM is defined with respect to a complete basis of two-electron eigenstates. For now we take this basis to be the two-particle Slater determinants
\begin{equation}\label{eq:nibasis}
	\psi_{\mathbf{n}}(\xx) = \frac{1}{\sqrt{2}}\left(\phi_{n_1}(\mathbf{x}_1)\phi_{n_2}(\mathbf{x}_2) - \phi_{n_1}(\mathbf{x}_2)\phi_{n_2}(\mathbf{x}_1)\right)
\end{equation} 
labeled by the pair of integers $\mathbf{n} = \conf{n_1,n_2}$. For a many-electron state defined by a single configuration $\conf{\alpha}$, we find in Appendix \ref{sec:Properties} that the $2$-RDM can be represented as a rank four tensor indexed by the pairs
\begin{equation}\label{eq:Drulenn}
	D_{\mathbf{mn}} = \begin{cases}
		1, & \text{if } \mathbf{m}=\mathbf{n}\text{ and } n_1,n_2 \in \conf{\alpha} \\
		0, & \text{otherwise}.
	\end{cases}
\end{equation}
To transform Eq. \ref{eq:Drulenn} into a matrix with indices $D_{mn}$, we must assign each pair $\mathbf{n}$ to a single integer. For this purpose we choose the mapping shown in Table \ref{tab:map}.
\begin{table}[h]
	\centering
	\caption{Geminal index map}
	\label{tab:map}
	\begin{tabular}[t]{lccccccc}
		\toprule
		$\mathbf{n}$ & $(1,2)$ & $(1,3)$ &$(2,3)$ & $(1,4)$ & $(2,4)$& $(3,4)$&\dots\\
		$n$ & $1$ & $2$ & $3$ & $4$ & $5$ & $6$ &$\dots$ \\
		\bottomrule
	\end{tabular}
\end{table}

We now have all the necessary tools to give an example; suppose our wave function is the three-electron Slater determinant $\Psi(\mathbf{x}_1,\mathbf{x}_2,\mathbf{x}_3) = \hat{S}_- \phi_1(\mathbf{x}_1)\phi_2(\mathbf{x}_2)\phi_3(\mathbf{x}_3)$. From Eq. \ref{eq:Drulenn} it is clear that $D_{\mathbf{nn}} = 1$ for $\mathbf{n} = \conf{1,2},\conf{1,3}$ and $\conf{2,3}$. Converting to a matrix by Table \ref{tab:map}, we find that $D_{nn} = 1$ for $n = 1,2,3$. This example generalizes trivially to $N$-electron Slater determinants, which are represented by matrices with $N(N-1)/2$ ones placed along the diagonal. Such matrices obey the strict equality of Eq. \ref{eq:tracesquared0}
\begin{equation}\label{eq:trtr1}
	\Tr[\mathbf{D}^2] = \nc
\end{equation}
and the equivalent idempotence condition
\begin{equation}\label{eq:DD1}
	\mathbf{D}^2 = \mathbf{D}.
\end{equation}

It is tempting to conclude that we can arbitrarily place ones on the diagonal, but this assumption fails immediately for the matrix $\mathbf{D}'$ with $D'_{nn} = 1$ for $n = 1,2,4$. According to Table \ref{tab:map}, the Slater determinant which generates this matrix must contains pairs $\conf{1,2},\conf{1,3},$ and $\conf{1,4}$.  It is impossible for a single three-electron product to contain all four basis functions $\phi_1(\mathbf{x}), \phi_2(\mathbf{x}), \phi_3(\mathbf{x})$ and $\phi_4(\mathbf{x})$, so the matrix is not $N$-representable. We conclude that the restrictions in Eq. \ref{eq:ruless} are insufficient to ensure $N$-representability because $\mathbf{D}'$ satisfies all the rules and fails to correspond to a valid wave function.

We may also create valid states that do not satisfy Eqs. \ref{eq:trtr1} and \ref{eq:DD1} by taking a linear superposition of configurations
\begin{equation}
	\Psi(\mathbf{x}_1,\dots,\mathbf{x}_N)= \sum_{\conf{\alpha}}C_{\conf{\alpha}}\Psi_{\conf{\alpha}}(\mathbf{x}_1,\dots,\mathbf{x}_N),
\end{equation}
whose diagonal elements are (Eq. \ref{eq:Dmnfinal})
\begin{align}\label{eq:Ddisj}
	D_{\mathbf{n}\mathbf{n}} = \sum_{\conf{\alpha}\ni \mathbf{n}}\left|C_{\conf{\alpha}}\right|^2.
\end{align}
The remaining elements $D_{\mathbf{mn}}$ with $\mathbf{m}\neq \mathbf{n}$ are only non-zero when two configurations share all but two basis functions (Eq. \ref{eq:Dndiag}). Take for example the constant superposition of $M$ disjoint anti-symmetrized configurations
\begin{align}\label{eq:Psidis}
	\Psi&(\mathbf{x}_1,\dots,\mathbf{x}_N) = \frac{1}{\sqrt{M}}\hat{S}_-\Big[\Big(\phi_1(\mathbf{x}_1)\cdots\phi_N(\mathbf{x}_N)\Big) \nonumber \\
	&+\Big(\phi_{N+1}(\mathbf{x}_1)\cdots\phi_{2N}(\mathbf{x}_N)\Big) \nonumber \\
	&+ \dots + \Big(\phi_{(M-1)N+1}(\mathbf{x}_1)\cdots\phi_{MN}(\mathbf{x}_N)\Big)\Big],
\end{align}
where each $\phi_i(\mathbf{x}) \neq \phi_j(\mathbf
x)$. Clearly, the off-diagonal elements of Eq. \ref{eq:Psidis} are all zero. We then find from Eq. \ref{eq:Ddisj} that $D_{\mathbf{nn}} = 1/M$ because each pair in the expansion appears in a single configuration with coefficient $1/\sqrt{M}$. It is straightforward to show that $\Tr[\mathbf{D}^2] = N(N-1)/(2M^2)$, which decreases as the number $M$ of disjoint configurations increases.

\subsection{Matrices representing a real system}\label{sec:real}
The previous examples were defined without regard to their relationship to a physical system. We now aim to make use of the GDM to solve the eigenvalue relation
\begin{equation}\label{eq:Hn}
	H(\mathbf{x}_1,\dots,\mathbf{x}_N)\Psi_i(\mathbf{x}_1,\dots,\mathbf{x}_N) = \EE_i \Psi_i(\mathbf{x}_1,\dots,\mathbf{x}_N)
\end{equation}
for an $N$-electron Hamiltonian $H(\mathbf{x}_1,\dots,\mathbf{x}_N)$ containing one and two-body terms as in Eq. \ref{eq:sop}. We discovered in Section \ref{sec:Density} that any wave function $\Psi_i(\mathbf{x}_1,\dots,\mathbf{x}_N)$ corresponds to a matrix $\mathbf{D}_i$ that produces the eigenvalues $\mathcal{E}_i$ of Eq. \ref{eq:Hn} by
\begin{equation}\label{eq:Eibasis}
	\EE_i = \Tr[\mathbf{D}_i\mathbf{H}].
\end{equation}
The matrix $\mathbf{H}$ has elements 
\begin{equation}\label{eq:Hmn}
	H_{mn} = \int \psi_m^*(\xx)H(\xx)\psi_n(\xx)d\xx,
\end{equation}
where $H(\xx)$ is the effective two-particle Hamiltonian (Eq. \ref{eq:A2}) acting on the geminal basis functions $\psi_i(\xx)$. Instead of using the two-electron Slater determinants of Eq. \ref{eq:nibasis}, it will be advantageous to choose the basis that diagonalizes $H(\xx)$ by
\begin{equation}\label{eq:H2basis}
H(\xx)\psi_j(\xx) = E_j\psi_j(\xx).
\end{equation}

Suppose that we have solved for at least the $N(N-1)/2$ lowest energy eigenstates of Eq. \ref{eq:Eibasis} and computed the diagonal matrix $\mathbf{H}$ by Eq. \ref{eq:Hmn}. We may then discover the ground state by finding the $N$-representable matrix $\mathbf{D}_i$ that yields the minimum possible energy in Eq. \ref{eq:Eibasis}. Dropping subscript $i$ from the GDM, the minimum energy state that satisfies the rules of Eq. \ref{eq:ruless} is
\begin{align}\label{eq:Drule}
	D_{mn} &= \begin{cases}
		1  & \text{if } m=n \text{ and } n \leq \nc \\
		0 & \text{otherwise}.
	\end{cases}
\end{align}
Unfortunately, the matrix in Eq. \ref{eq:Drule} is not guaranteed to be $N$-representable because it was discovered through a minimization subject to an insufficient set of constraints. It is nonetheless worthwhile to introduce Eq. \ref{eq:Drule} for two reasons, the first being that it is precisely the form of the ground state matrices postulated by Bopp \cite{boppAbleitungBindungsenergieNTeilchenSystemen1959} for his atomic calculations. The second use is to gain insight into the $N$-representability problem and the path toward its resolution.

We may wonder if the correct ground state GDM may be non-diagonal in contrast to Eq. \ref{eq:Drule}. However, borrowing intuition from single-particle mixed-case density matrices, we expect off-diagonal elements to introduce temporal density oscillations that render the state non-stationary. Another alternative is that the eigenstate matrices are diagonal but the elements may be any real number between 0 and 1. In this case, the GDM formalism seems not to reduce difficulty of the energy minimization because the optimization occurs over a possibly infinite set of matrix coefficients.

There is a hint following from a physical argument that stationary states do not exhibit such non-integral occupation levels. Suppose $N$ electrons begin in a Slater determinant in a system governed by a Hamiltonian with electron-electron interactions. The initial GDM is trivially $N$-representable as it is constructed directly from the wave function, but it is clearly not stationary. Given enough time, however, we expect the system to relax to the ground state by processes like the emission of radiation. Using single particle density matrices as a guide again, we may conjecture that the electron-radiation interaction proceeds as a unitary transformation of the GDM.

Since $\Tr[\mathbf{D}^2]$ is conserved under unitary transformation, the ground state must also satisfy $\Tr[\mathbf{D}^2] = N(N-1)/2$ and $\mathbf{D}^2 = \mathbf{D}$. These conditions are inconsistent with the presence of non-integer occupation numbers that decrease the trace of the squared GDM. Accordingly, we may predict that all stationary states are represented by $N(N-1)/2$ ones along the diagonal but the remaining difficulty is to determine which collection of occupied states defines an $N$-representable GDM.

The preceding discussion suggests that Bopp's model may be closer to correct than previously thought. After all, his method calculated the ground state of O$^{5+}$ with a remarkably low error of $0.017\%$ (See tables I and II of Ref \cite{boppAbleitungBindungsenergieNTeilchenSystemen1959}). Although the method fared worse for Be$^+$ with an error of $0.86\%$, the matrix designated as the first excited state curiously had energy within $0.040\%$ of the experimental ground state. While it is always possible that these anomalies are the result of coincidence, it deserves to be investigated whether Bopp's errors originated from a simple improper accounting of two-electron eigenstates.

Clearly, it is not immediately obvious that the GDM should actually evolve by unitary matrix transformation. For this reason, the next section carefully derives the evolution law. 

\section{The time evolution equation}\label{sec:Time}
Subsection \ref{sec:real} emphasized the importance of modeling the time evolution of the GDM for the discovery of many-body eigenstates. In this section we derive the governing equation, finding that the GDM indeed evolves in unitary fashion by the Liouville-Von Neumann equation.

We begin by generalizing the equations of Section \ref{sec:Density} to apply at arbitrary times. Most importantly, our wave function will be the position representation of the time-dependent abstract operator $\alpha(t)$
\begin{equation}\label{eq:psiabs}
	\Psi(\mathbf{x}_1,\dots,\mathbf{x}_N|t) = \braket{\mathbf{x}_1,\dots,\mathbf{x}_N|\alpha(t)}.
\end{equation}
The expectation value of a generally time-dependent linear operator $\hat{A}(t)$ for the state in Eq. \ref{eq:psiabs} is
\begin{equation}\label{eq:Awave}
	\braket{A}(t) = \braket{\alpha(t)|\hat{A}(t)|\alpha(t)}.
\end{equation}
We can also calculate observables by the equivalent $2$-RDM formulation, defining the time-dependent $2$-RDM
\begin{equation}\label{eq:rt}
	\rho(\xx,\xx'|t) = \int \Psi^*(\xx,\yy|t)\Psi(\xx',\yy|t)d\yy
\end{equation}
following Eq. \ref{eq:rho1}. The expectation value is then given by
\begin{equation}\label{eq:rhoev}
	\braket{A}(t) = \int \delta(\xx-\xx') A(\xx'|t)\rho(\xx,\xx'|t)d\xx d\xx'.
\end{equation}
The question that we need to answer is: How must $\rho(\xx,\xx'|t)$ evolve in time so that the expectation values computed by Eq. \ref{eq:rhoev} match those given in Eq. \ref{eq:Awave} by the many-body wave function? The ability to compute the same observable quantities for a given time means that $\rho(\xx,\xx'|t)$ and its corresponding $\mathbf{D}(t)$ furnish a complete representation of the quantum state. 

The first step in answering the posed question is to differentiate Eq. \ref{eq:rhoev} to find
\begin{align}\label{eq:rhocomp}
	\frac{d}{dt}\braket{A}(t) &=  \Braket{\frac{dA}{dt}} \nonumber \\ 
	&+ \int  d\xx d\xx'\delta(\xx - \xx')A(\xx')\dot{\rho}(\xx,\xx'|t).
\end{align}
We must find a matching expression for Eq. \ref{eq:rt} in the hope of finding an equation that connects $\dot{\rho}(\xx,\xx'|t)$ to known quantities. For brevity, we will group all $N$ coordinate sets together into the symbol 
\begin{equation}
	\overline{\xx} = \mathbf{x}_1,\dots,\mathbf{x}_N
\end{equation}
so that we can express the $N$-electron wave function at some initial time $t_0$ as $\Psi(\overline{\xx}|t_0)$. This wave function evolves according to the unitary time evolution operator $U(t,t_0)$ by 
\begin{equation}
	\Psi(\overline{\xx}|t) = U(t,t_0)\Psi(\overline{\xx}|t_0).
\end{equation}
Expressing the time-dependent operator $A(\overline{\xx}|t)$ as a pair sum following Eq. \ref{eq:Atot}, we find the position representation of the inner product in Eq. \ref{eq:Awave}
\begin{align}\label{eq:At0}
    \braket{A}(t) &= \sum_{i,j>i}\int \Psi^*(\overline{\xx}|t_0)U^\dagger(t,t_0) \nonumber \\
     &\times A(\mathbf{x}_i,\mathbf{x}_j|t) U(t,t_0) \Psi(\overline{\xx}|t_0)d\overline{\xx}.
\end{align}
Recalling the notation of Section \ref{sec:Density}, we have defined the differential $d\overline{\xx} = d\mathbf{x}_1\cdots d\mathbf{x}_N$ which includes spin sums by Eq. \ref{eq:spinsum}.

We proceed to differentiate Eq. \ref{eq:At0}, distributing derivatives by the chain rule to $U^\dagger(t,t_0)$, $A(\mathbf{x}_i,\mathbf{x}_j)$ and $U(t,t_0)$. Applying the time-dependent Schr{\"o}dinger equations $i\partial_t U(t) = H(t) U(t)$ and $-i\partial_t U^\dagger(t) = U^\dagger(t) H(t)$, we find that
\begin{align}\label{eq:Along1}
	\frac{d}{dt}\braket{A}(t) = \Braket{\frac{dA}{dt}} -iK(t),
\end{align}
where $K(t)$ is the expectation value of the many-body commutator
\begin{align}\label{eq:Along2}
	K(t) &=  \int d\overline{\xx}\Psi^*(\overline{\xx}|t) \nonumber \\
	&\times\sum_{\substack{i,j>i \\ k,l>k}}
	\left[H(\mathbf{x}_i,\mathbf{x}_j|t),A(\mathbf{x}_k,\mathbf{x}_l|t)\right]\Psi(\overline{\xx}|t)d\overline{\xx}.
\end{align}
In Eq. \ref{eq:Along2} we expressed the many-body Hamiltonian $H(\overline{\xx}|t)$ as a pair sum per Eq. \ref{eq:Atot}. 

Simplify Eq. \ref{eq:Along2} by splitting $H(\mathbf{x}_i,\mathbf{x}_j|t)$ and $A(\mathbf{x}_k,\mathbf{x}_l|t)$ into their one and two body components following Eq. \ref{eq:Asep1}. Defining $K_{\alpha \beta}(t)$ to be the integral involving the commutator between $\alpha$-body terms of the Hamiltonian and $\beta$-body terms of $A$, $K(t)$ decomposes into the sum
\begin{equation}\label{eq:Ktot}
	K(t) = K_{11}(t) + K_{12}(t) + K_{21}(t) + K_{22}(t).
\end{equation}

We will now calculate $K_{1\beta}(t)$, which deals with the commutator between one-body Hamiltonian terms and both one and two-body portions of $A$. By swapping each $\mathbf{x}_k$ to $\mathbf{x}_1$ and $\mathbf{x}_l$ to $\mathbf{x}_2$, the sum over $k$ and $l>k$ in \ref{eq:Along2} reduces to $N(N-1)/2$ identical integrals. Continuing to expand $H_1(\mathbf{x}_i,\mathbf{x}_j)$ by Eq. \ref{eq:Aeff1}, we can separate $\mathbf{x}_i$ and $\mathbf{x}_j$ contributions to find
\begin{align}\label{eq:K1b}
	K_{1\beta}(t) &= \nc \int d\overline{\xx} \Psi^*(\overline{\xx}|t) \nonumber \\
	\times\Bigg(&\sum_{i,j>i}\left[\frac{h_1(\mathbf{x}_i|t)}{N-1},A_{\beta}(\mathbf{x}_1,\mathbf{x}_2|t)\right] \nonumber \\
	+&\sum_{i,j>i}\left[\frac{h_1(\mathbf{x}_j|t)}{N-1},A_{\beta}(\mathbf{x}_1,\mathbf{x}_2|t)\right]\Bigg)\Psi(\overline{\xx}|t).
\end{align}

Most terms in Eq. \ref{eq:K1b} cancel because operators acting on different coordinates always commute. The only surviving contributions are those in which $\mathbf{x}_i$ or $\mathbf{x}_j$ are equal to $\mathbf{x}_1$ or $\mathbf{x}_2$. From the second line we pick up $N-1$ copies of the commutator $[h_1(\mathbf{x}_1), A_\beta(\mathbf{x}_1,\mathbf{x}_2)]$ by fixing $i=1$ and running over all $N-1$ elements in the $j$ sum. We also find from this expression a single instance of the commutator  $[h_1(\mathbf{x}_2), A_\beta(\mathbf{x}_1,\mathbf{x}_2)]$. In the third line we find $N-2$ more copies of the commutator $[h_1(\mathbf{x}_2), A_\beta(\mathbf{x}_1,\mathbf{x}_2)]$ by fixing $i=2$ and summing over the $N-2$ values of $j$, so that each portion appears a total of $N-1$ times.

Noting that the above argument applies identically to the computation of $K_{\alpha 1}$, we find that
\begin{align}\label{eq:Kab}
	K_{\alpha \beta}(t) &= \nc (N-1)\int  \Psi^*(\overline{\xx}|t) \nonumber \\
	&\times\left[H_{\alpha}(\xx|t),A_{\beta}(\xx|t)\right]\Psi(\overline{\xx}|t) d\overline{\xx}
\end{align}
for $\alpha \beta = 11,12,21$. Eq. \ref{eq:Kab} does not immediately apply to $K_{22}(t)$ because, unlike in Eq. \ref{eq:K1b}, two-body operators cannot be separated into single-coordinate expressions. As a result, 3-coordinate terms such as $[H(\mathbf{x}_1,\mathbf{x}_2), A(\mathbf{x}_1,\mathbf{x}_3)]$ remain in the equation for $K_{22}(t)$. By restricting two-body terms of the Hamiltonian and operator $A$ to depend only on position in the form $\sum_{i,j>i}f(|\mathbf{r}_i - \mathbf{r}_j||t)$, two body terms will always commute so that $[H_2(\xx|t),A_2(\xx|t)] = 0$ and Eq. \ref{eq:Kab} applies trivially to $\alpha \beta = 22$. The consequences of this requirement are discussed in the Conclusion.

Absorb the prefactor $(N-1)$ of Eq. \ref{eq:Kab} into the Hamiltonian by defining
\begin{equation}\label{eq:Hprime}
	H'(\xx|t) = (N-1)H(\xx|t).
\end{equation}
Summing the $K_{\alpha\beta}$ by Eq. \ref{eq:Ktot} allows us to express $K(t)$ as
\begin{align}\label{eq:Kwf}
	K(t) &=  \nc \int d\xx d\yy \Big\{ \nonumber \\
	&  \Big[H'(\xx|t)\Psi^*(\xx,\yy|t)\Big] 
	A(\xx|t)\Psi(\xx,\yy|t) \nonumber \\
	&- \Psi^*(\xx,\yy|t)H'(\xx|t) 
	A(\xx|t)\Psi(\xx,\yy|t)\Big\}.
\end{align}
In Eq. \ref{eq:Kwf}, we separated the terms of the commutator and chose one Hermitian operator $H'(\xx|t)$ to act on the left copy of the wave function. We can finally substitute Eq. \ref{eq:rt} into Eq. \ref{eq:Kwf} to find the derivative of the expectation value by Eq. \ref{eq:Along1}:
\begin{align}\label{eq:Afromr}
	\frac{d}{dt}\braket{A}(t) &= \Braket{\frac{dA}{dt}}  -i\int d\xx d\xx' \delta(\xx - \xx') \\
	&\times A(\xx')[H'(\xx) - H'(\xx')]\rho(\xx,\xx'|t).
\end{align}
Note that one copy of the Hamiltonian depends on coordinates $\xx$ as a consequence of its acting to the left in the inner product. Comparing Eq. \ref{eq:Afromr} to Eq. \ref{eq:rhocomp}, we find the desired expression for $\dot{\rho}(\xx,\xx'|t)$:
\begin{equation}\label{eq:rhodot}
	\dot{\rho}(\xx,\xx'|t) = -i[H(\xx) - H(\xx')]\rho(\xx,\xx'|t).
\end{equation}

We proceed to derive a more convenient matrix representation of Eq. \ref{eq:rhodot} starting with a time-dependent geminal expansion of the $2$-RDM
\begin{equation}\label{eq:rhoegen}
	\rho(\xx,\xx'|t) = \sum_{mn}D_{mn}(t) \psi_n^*(\xx|t)\psi_m(\xx'|t).
\end{equation}
Note that Eq. \ref{eq:rhoegen} represents the most general case in which both the matrix elements $D_{mn}(t)$ and the geminal basis functions may vary in time. Plugging into Eq. \ref{eq:rhodot} gives
\begin{align}\label{eq:derivative}
	i\frac{d}{dt}&\braket{\xx'|\hat{D}|\xx} =  \sum_{mn}D_{mn}(t) \psi_m(\xx'|t)\left[H'(\xx|t)\psi_n^*(\xx|t)\right] \nonumber \\
	&- \sum_{mn}D_{mn}(t) \left[H'(\xx'|t)\psi_m(\xx'|t)\right]\psi_n^*(\xx|t),
\end{align}
where we have rearranged the $\psi_i(\xx)$ into the most convenient order. Substitute into Eq. \ref{eq:derivative} the following identities
\begin{align}\label{eq:Dcom}
	\psi_m(\xx'|t) &= \braket{\xx'|m(t)} \\
	\psi_n^*(\xx|t) &= \braket{n(t)|\xx} \\
	H'(\xx'|t)\psi_m(\xx'|t) &= \braket{\xx'|\hat{H}'(t)|m(t)} \\
	H'(\xx|t)\psi_{n}^*(\xx|t) &= \braket{n(t)|\hat{H}'(t)|\xx},
\end{align}
so that everything on the right-hand side sits between $\bra{\xx'}$ and $\ket{\xx}$. Recalling the definition of $\hat{D}(t)$ (Eq. \ref{eq:Dab}), we extract from Eq. \ref{eq:derivative} the abstract operator equation 
\begin{equation}\label{eq:comabs}
	\frac{d}{dt}\hat{D}(t) = -i[\hat{H}'(t),\hat{D}(t)].
\end{equation}
Eq. \ref{eq:comabs} is the familiar Liouville-Von Neumann equation.

We can specialize Eq. \ref{eq:comabs} to a matrix equation by choosing a time-independent basis $\ket{i}$ with which to expand $\hat{D}(t) = \sum_{mn}D_{mn}\ket{m}\bra{n}$ and $\hat{H}(t) = H_{mn}\ket{m}\bra{n}$. The result is that
\begin{equation}\label{eq:Lvn}
    \dot{\mathbf{D}}(t) = -i[\mathbf{H}'(t),\mathbf{D}(t)],
\end{equation} 
which we find by an alternate derivation in Appendix \ref{sec:Alternate}. Eq. \ref{eq:Lvn} can be shown to evolve through matrix transformation by the operator $\mathbf{U}(t,t_0)$ as
\begin{equation}\label{eq:Dopu}
    \mathbf{D}(t) = \mathbf{U}(t,t_0)\mathbf{D}(t_0)\mathbf{U}^\dagger(t,t_0).
\end{equation}
Plugging Eq. \ref{eq:Dopu} into Eq. \ref{eq:Lvn} and matching expressions on either side we find that $\mathbf{U}(t)$ must satisfy
\begin{equation}\label{eq:UUU}
	\dot{\mathbf{U}}(t,t_0) = -i \mathbf{H}(t)\mathbf{U}(t,t_0),
\end{equation}
and it is a routine procedure to successively integrate Eq. \ref{eq:UUU} to infinite order for
\begin{equation}\label{eq:U}
    \mathbf{U}(t,t_0) = \mathcal{T}_t \exp\left(-i\int_{t_0}^t d\tau \mathbf{H}'(\tau)\right).
\end{equation}
$\mathbf{U}(t)$, being the exponential of a skew-Hermitian matrix, is clearly unitary. As mentioned at the end of Section \ref{sec:Density}, we now see that the unitary transformation-invariant property
\begin{equation}\label{eq:idemp}
	\mathbf{D}^2 = \mathbf{D}
\end{equation}
holds for any state connected to a Slater determinant by a time-dependent Hamiltonian. Since states can only change by such a unitary transformation, Eq. \ref{eq:idemp} must be true for any GDM with fixed particle number.

\section{Switching on the electron-electron interaction}\label{sec:Evolution}
Having confirmed unitary evolution of the GDM, we will now exploit this property to solve the many-body Schr\"odinger equation. We do so by studying an $N$-electron Hamiltonian in which the Coulomb potential is switched on by a temporal function $\lambda(t)$. Using Hartree atomic units with $\hbar = m_0 = |e|= 1/(4\pi \epsilon_0) = 1$, the Hamiltonian is written
\begin{equation}\label{eq:Hamfull}
	H(\overline{\xx}|t) = \sum_{i,j>i} H_1(\mathbf{x}_i,\mathbf{x}_j) + \frac{\lambda(t)}{|\mathbf{r}_i - \mathbf{r}_j|},
\end{equation}
with $\lambda(0) = 0 $ and $\lambda(t) = 1$ for $t$ larger than some switching time $T$. 

As we saw in Eq. \ref{eq:A2}, Eq. \ref{eq:Hamfull} reduces in the $2$-RDM formalism to the effective two-electron form
\begin{equation}\label{eq:Ht}
	H(\xx|t) = H_1(\xx) + \frac{\lambda(t)}{|\mathbf{r}_1 - \mathbf{r}_2|}.
\end{equation}

We introduce two natural basis sets for the study of Eq. \ref{eq:Ht}, the first being the non-interacting basis that  diagonalizes the Hamiltonian at $t=0$ by
\begin{equation}
	H(\xx|0)\psi_i(\xx) = E_i \psi_i(\xx).
\end{equation}
Similarly, we define the interacting basis by fixing $t=T$ and solving
\begin{align}
	H(\xx|T)\psi_i(\xx) &= \left(H_1(\xx) + \frac{1}{|\mathbf{r}_1 - \mathbf{r}_2|}\right)\psi_{i}(\xx) \nonumber \\
	&= E_i \psi_i(\xx).
\end{align}
Using the subscript $I$ and $N$ for the interacting and non-interacting bases, respectively, we can change between the two representations by the unitary transformation (see Appendix \ref{sec:Basis}):
\begin{equation}\label{eq:baschange}
	\mathbf{D}_I(t) = \mathbf{U}_{I}^N\mathbf{D}_N(t)\big(\mathbf{U}_{I}^N\big)^\dagger.
\end{equation}

In the following, we observe the time evolution of a Slater determinant under the influence of the Hamiltonian in Eqs. \ref{eq:Hamfull} and \ref{eq:Ht}. Section \ref{sec:sudden} treats the case in which the Coulomb interaction is quickly switched on and Section \ref{sec:adiabatic} details a slow adiabatic change which allows us to construct fully-interacting many-body solutions.

\subsection{The sudden approximation}\label{sec:sudden}

Beginning with the electrons in a Slater determinant state, we instantaneously turn on the Coulomb interaction by the step function $\lambda(t) = u(t-T)$. The sudden approximation posits that the electron gas remains unchanged during switching but starts to evolve according to the Hamiltonian $H(\overline{\xx}|t)$ for $t>T$. Since the Hamiltonian is constant for $t>T$, the time-dependent GDM is found from Eqs. \ref{eq:Dopu} and \ref{eq:U} to be
\begin{equation}\label{eq:bt}
	\mathbf{D}(t) = e^{i\mathbf{H}'(T)(t-T)}\mathbf{D}(T)e^{-i\mathbf{H}'(T)(t-T)}.
\end{equation}
Selecting the interacting basis, we have that the Hamiltonian $\mathbf{H}$ is diagonal while $\mathbf{D}(T)$ is non-diagonal by the transformation in Eq. \ref{eq:baschange}. This choice of basis simplifies the matrix equation in Eq. \ref{eq:bt} to
\begin{equation}\label{eq:Dtt}
	D_{mn}(t) = e^{iE_{nm}'(t-T)}D_{mn}(T),
\end{equation}
with $E'_{mn} = (N-1)(E_m - E_n)$ accounting for the multiplicative constant attached to $H'(\xx)$ in Eq. \ref{eq:Hprime}.

With the time-dependent GDM given by Eq. \ref{eq:Dtt}, we can calculate the observable quantity described by the operator with position representation
\begin{equation}\label{eq:density}
	\rho(\mathbf{x};\xx) = \frac{\delta(\mathbf{x}_1 - \mathbf{x}) + \delta(\mathbf{x}_2 - \mathbf{x})}{N-1}.
\end{equation}
Recalling that $\mathbf{x} = \mathbf{r}\sigma$, the expectation value $\braket{\rho(\mathbf{x})}(t)$ gives the electron density at position $\mathbf{r}$ with spin $\sigma$. For simplicity, we will define the symbol
\begin{equation}
	\rho(\mathbf{x},t) = \braket{\rho(\mathbf{x})}(t),
\end{equation} 
which we calculate by the trace relation
\begin{equation}\label{eq:rhotrace}
	\rho(\mathbf{x},t) = \Tr[\mathbf{D}(t)\boldsymbol{\rho}(\mathbf{x})].
\end{equation}
The matrix elements of $\boldsymbol{\rho}(\mathbf{x})$ take the simplified form
\begin{equation}\label{eq:rhot2}
	\rho_{nm}(\mathbf{x}) = \frac{2}{N-1}\int \psi_{n}^*(\mathbf{x},\mathbf{x}_2)\psi_{m}(\mathbf{x},\mathbf{x}_2)d\mathbf{x}_2.
\end{equation}

Computing the trace in Eq. \ref{eq:rhotrace} finally yields the electron density
\begin{equation}\label{eq:rhot}
    \rho(\mathbf{x},t) = \sum_{mn}D_{mn}(T)e^{-iE'_{mn}(t-T)}\rho_{nm}(\mathbf{x}).
\end{equation}
We perform a simple check that particle number is conserved; integrating Eq. \ref{eq:rhot} over $\mathbf{x}$, we find from orthonormality and Eq. \ref{eq:trace0} that $ \int \rho(\mathbf{x},t) d\mathbf{x} = N$ as required. In the next subsection we delve deeper into the implications of the electron density equation.
\subsubsection{The origin of incoherent quantum fluctuations}
We can extract a surprising amount of insight from the simple relation in Eq. \ref{eq:rhot}. Separating the stationary and oscillating terms allows us to express the density as
\begin{align}\label{eq:rhoo}
    \rho(\mathbf{x},t) &= \sum_l D_{ll}\rho_{ll}(\mathbf{x}) + \sum_{\substack{m,n\\E_{m}'=E_n'}}\text{Re}\{P_{mn}(\mathbf{x})\} \nonumber \\
    &+ \sum_{\substack{p,q\\E_{p}'\neq E_q'}}|P_{pq}(\mathbf{x})|\cos\left(E_{pq}'t + \theta_{pq}(\mathbf{x})\right),
\end{align}
with 
\begin{align}
	P_{ij}(\mathbf{x}) &= D_{ij}(T)\rho_{ji}(\mathbf{x}) \\
		\theta_{ij}(\mathbf{x}) &= \arg \left( P_{ij}(\mathbf{x})\right).
\end{align}
The first sum in Eq. \ref{eq:rhoo} gives the contribution to the density from diagonal matrix elements, and the second and third are from the degenerate and non-degenerate off-diagonals, respectively.

Defining the time average functional for $t > T$ by
\begin{equation}
	\overline{f(\mathbf{x},t)} = \lim_{\tau \rightarrow \infty}\frac{1}{\tau}\int_{T}^{T+\tau}f(\mathbf{x},t')dt',
\end{equation}
we find the time-averaged density at $\mathbf{x}$ to be determined entirely by the diagonal and degenerate terms
\begin{equation}
	\overline{\rho(\mathbf{x},t)} = \sum_l D_{ll}\rho_{ll}(\mathbf{x}) + \sum_{\substack{m,n\\E_{m}'=E_n'}}\text{Re}\{P_{mn}(\mathbf{x})\}.
\end{equation}
On the other hand, the non-degenerate off-diagonals introduce temporal density fluctuations with variance
\begin{equation}\label{eq:sigma}
    \sigma^2(\mathbf{x}) = \overline{\rho^2(\mathbf{x},t)} - \overline{\rho(\mathbf{x},t)}^2  = \frac{1}{2}\sum_{\substack{p,q\\E_{p}'\neq E_q'}}|P_{pq}(\mathbf{x})|^2.
\end{equation}
The fluctuations increase in magnitude as the number of off-diagonal terms increases, an idea which can also be understood directly from Eq. \ref{eq:rhoo} whereby summing an increasing number of out-of-phase cosines leads to peaks in the density that decrease in duration but increase in intensity. A similar argument can be made for the spatial extent of these quasi-random density spikes.

Eqs. \ref{eq:rhoo}--\ref{eq:sigma} are best understood in the context of a physical example. Suppose we have a system of electrons experiencing the potential of a set of nuclei at positions $\mathbf{R}_i(t)$ that are free to move in time. The resulting effective electronic Hamiltonian is
\begin{align}\label{eq:Hpr}
	H'(\xx|t) &= \sum_{i=1}^2\left(-\frac{\nabla_i^2}{2} - \sum_j\frac{1}{|\mathbf{r}_i - \mathbf{R}_j(t)|}\right) \nonumber \\
	&+ (N-1)\frac{\lambda(t)}{|\mathbf{r}_1 - \mathbf{r}_2|},
\end{align}
noting that we used $H'(\xx|t) = (N-1)H(\xx|t)$. Once again the presence of $\lambda(t) = u(t-T)$ indicates that we abruptly switch on the Coulomb potential at time $t=T$. 

The electron-nuclei system evolves according to the coupled equations 
\begin{align}
	\dot{\mathbf{D}}(t) &= -i[\mathbf{H}'(t),\mathbf{D}(t)]\label{eq:Dcou} \\
	M_i\ddot{\mathbf{R}}_i(t) &= \sum_{j\neq i}\frac{1}{|\mathbf{R}_i(t) - \mathbf{R}_j(t)|^2} - \sum_{\sigma}\int \frac{\rho(\mathbf{r}\sigma,t)}{|\mathbf{r} - \mathbf{R}_i|^2}d\mathbf{r} \label{eq:Mc},
\end{align}
where Eq. \ref{eq:Mc} is the classical non-relativistic equation of motion for nucleus $i$ with mass $M_i$. Each nucleus feels a repulsive Coulomb force induced by the other nuclei and an attractive Coulomb force from the electron gas. The sum over electrons spins is explicitly written following Eq. \ref{eq:spinsum}. We take the system to be at rest for $t<T$, meaning that all $\mathbf{R}_i(t)$ are fixed in place and the electrons are in an eigenstate represented by a diagonal matrix. 

When $t>T$, the electron density fluctuates by a series of spikes localized in space and time. Each peak causes an abrupt change in the electron-nucleus Coulomb force in Eq. \ref{eq:Mc} which causes a near-instantaneous scattering of the nuclei. The resulting motion of $\mathbf{R}_i(t)$ affects the Hamiltonian in Eq. \ref{eq:Hpr} and subsequently the GDM by Eq. \ref{eq:Dcou}. The net effect is the transfer of energy from the electrons to the nuclei. This process continues until an equilibrium is reached wherein net energy ceases to flow between the subsystems. There is an alternate picture that expands Eq. \ref{eq:Mc} as a sum of normal modes of motion excited by the oscillatory terms of Eq. \ref{eq:rhoo}. Summing the out-of-phase normal mode oscillation amplitudes yields the same thermalized motion as the scattering picture.

Instead of evolving by the non-physical Hamiltonian in Eq. \ref{eq:Hpr}, we could have fixed $\lambda = 1$ and included the effect of a time dependent electromagnetic potential $A(\mathbf{r},t)$. In this case we would see that the field excites electron density fluctuations by introducing non-zero off-diagonal elements to $\mathbf{D}$, which induces motion in the nuclei leading again to thermalization. This example and the previous one illustrate a major strength of the GDM formalism. By properly treating all electrons together, we begin to see the emergence of classical behavior in a quantum system.

\subsection{The degenerate adiabatic theorem}\label{sec:adiabatic}
Section \ref{sec:Time} derived the equation of motion (Eq. \ref{eq:comabs}) 
\begin{equation}\label{eq:em2}
	\frac{d}{dt}\hat{D}(t) = -i[\hat{H}'(t),\hat{D}(t)]
\end{equation}
for the abstract density operator $\hat{D}(t)$. We found the corresponding matrix equation (Eq. \ref{eq:Lvn}) after expanding $\hat{D}(t)$ in a time-independent geminal basis. 

Here we derive the adiabatic theorem, beginning by determining the matrix equation of motion for a GDM expanded in a time-dependent basis in which the Hamiltonian is always diagonal. That is, the basis functions satisfy
\begin{equation}\label{eq:evec}
	\hat{H}(t)\ket{i(t)} = E_i(t)\ket{i(t)},
\end{equation}
where $E_i(t)$ is the instantaneous eigenvalue of state $\ket{i(t)}$ at time $t$. Labeling matrix elements by $\mathcal{D}_{pq}(t)$, we have that
\begin{equation}\label{eq:adex}
	\hat{D}(t) = \sum_{pq}\DD_{pq}(t)\ket{p(t)}\bra{q(t)}.
\end{equation}
Inserting Eq. \ref{eq:adex} into Eq. \ref{eq:em2} and picking out the $mn$ component by pre-multiplying $\bra{m(t)}$ and post-multiplying $\ket{n(t)}$, we find that the right-hand side evaluates to $-iE_{mn}'(t)D_{mn}(t)$ after applying Eq. \ref{eq:evec}. Repeating for the left-hand side, by the chain rule we have that
\begin{align}\label{eq:emb}
	&\Braket{m(t)|\frac{d}{dt}\hat{D}(t)|n(t)} = \dot{\DD}_{mn}(t) \nonumber \\ 
	&+\sum_{p}\mathcal{D}_{pn}(t)\braket{m(t)|\dot{p}(t)} + \sum_{q}\mathcal{D}_{mq}(t)\braket{\dot{q}(t)|n(t)}.
\end{align}
Because $(d/dt)\braket{q(t)|n(t)} = 0$ by orthonormality at time $t$, we can simplify Eq. \ref{eq:emb} by substituting $\braket{\dot{q}(t)|n(t)} \rightarrow -\braket{q(t)|\dot{n}(t)}$ and taking these terms to be the elements of the skew-Hermitian matrix $\mathbf{M}$ with coefficients
\begin{equation}\label{eq:Mel}
	M_{ij}(t) = \Braket{i(t)|\frac{d}{dt}|j(t)}.
\end{equation}
After doing so, the final equation of motion for the $mn$ element is
\begin{align}\label{eq:Dmnexpansion}
	&\dot{\mathcal{D}}_{mn}(t) = -iE_{mn}'(t)\mathcal{D}_{mn}(t) \nonumber \\
	 &-\sum_{p}\mathcal{D}_{pn}(t)M_{mp}(t) + \sum_{q}\mathcal{D}_{mq}(t)M_{qn}(t).
\end{align}

We will now simplify Eq. \ref{eq:Dmnexpansion} for a Hamiltonian that varies slowly over a long time interval $T$. In the context of this work, $T$ will be the switching time that appears in the Coulomb potential scaling $\lambda(t)$ of Eq. \ref{eq:Ht}. Closely following Ref. \cite{aguiarpintoAdiabaticApproximationDensity2002}, we will discover how $\boldsymbol{\mathcal{D}}(t)$ evolves when $T \rightarrow \infty$. Defining a natural time
\begin{equation}
	s(t) = \frac{t}{T}
\end{equation}
scaled by the switching duration, we use the fact that $(d/dt) = T(d/ds)$ to see that
\begin{align}\label{eq:qqq}
	\dot{\DD}_{mn}(s) &= - iTE_{mn}'(s) \DD_{pn}(s) \nonumber \\
	+&\sum_{q}\DD_{mq}(s)M_{qn}(s) - \sum_{p}M_{mp}(s)\DD_{pn}(s).
\end{align}
In Eq. \ref{eq:qqq} we used that $M_{ij}(s) = (1/T)M_{ij}(t)$ and chose a more convenient ordering for the sum terms. The first term on the right hand side represents the dynamical phase, which we factor out by defining $\tilde{\DD}(t)$ as
\begin{equation}\label{eq:dynphase}
	\DD_{mn}(s) =  \tilde{\DD}_{mn}(s)e^{-iT\int_{0}^s E_{mn}'(s') ds'},
\end{equation}
so that
\begin{align}\label{eq:Dtilde}
	&\frac{d}{ds}\tilde{\DD}_{mn}(s) = \hspace{-4pt}\sum_{\substack{q\\E_q=E_n}}\hspace{-4pt}\tilde{\DD}_{mq}(s)M_{qn}(s) -\hspace{-4pt}\sum_{\substack{p\\E_p=E_m}}\hspace{-4pt}M_{mp}(s)\tilde{\DD}_{pn}(s) \nonumber \\
	&+ \sum_{\substack{q\\E_q\neq E_n}}\frac{d}{ds}\int_0^s ds' \tilde{\DD}_{mq}(s')M_{qn}(s')e^{-iT\int_0^{s'} E_{nq}'(s'')ds''} \nonumber \\
	&-\sum_{\substack{p\\E_p\neq E_m}} \frac{d}{ds}\int_0^s ds' M_{mp}(s')\tilde{\DD}_{pn}(s')e^{-iT\int_0^{s'} E_{pm}'(s'')ds''}.
\end{align}
In Eq. \ref{eq:Dtilde} we separated the $p$ and $q$ sums into terms that accumulate dynamical phase and those that do not. To the latter we have applied the identity operator $\hat{1} = (d/ds)\int_0^s ds'$.

We will now evaluate the last line of Eq. \ref{eq:Dtilde} to understand the implications of the dynamical phase. To begin, we simplify by defining
\begin{equation}\label{eq:Fp}
	F_p(s') = M_{mp}(s')\DD_{pn}(s')
\end{equation}
and 
\begin{equation}\label{eq:gmp}
	g_{pm}(s') = \int_0^{s'} E_{pm}'(s'') ds''
\end{equation}
so that the integral over $ds'$ (which we call $I_p(s)$) takes the form
\begin{equation}
	I_{p}(s) = \int_{0}^s  F_p(s')e^{-iT g_{pm}(s')}ds'.
\end{equation}
Using that $\dot{g}_{pm}(s') = E'_{pm}(s')$ by Eq. \ref{eq:gmp}, we multiply the integrand by the identity $1 = iT\dot{g}_{pm}(s')/(iTE'_{pm}(s'))$ so that
\begin{equation}\label{eq:Ipa}
	I_{p}(s) = \int_{0}^s  \frac{F_p(s')}{iTE_{pm}'(s')}\left(iT\dot{g}_{pm}(s')e^{iT g_{pm}(s')}\right)ds'.
\end{equation}
Because $iT \dot{g}_{pm}(s')e^{iT g_{pm}(s')} = (d/ds')e^{iT g_{pm}(s')}$, we are able to integrate Eq. \ref{eq:Ipa} parts to finally find
\begin{align}\label{eq:Ip}
	I_p(s) &= \frac{1}{iT}\left[\frac{F_{p}(s')}{E_{pm}'(s')}e^{iTg_{pm}(s')}\right]_{0}^s \nonumber \\
	&- \frac{1}{iT}\int_0^s \frac{d}{ds'}\left(\frac{F_{p}(s')}{E'_{pm}(s')}\right)e^{iTg_{pm}(s')}ds'.
\end{align}
As long as $F_p(s')$ is differentiable, we can take the adiabatic approximation by letting $T\rightarrow \infty$ so that $I_p(s) \rightarrow 0$. Since the integral in the second line proceeds identically, we conclude that any contribution that accumulates dynamical phase will evaluate to zero.

Canceling these integral terms and reverting to the unscaled time $t$ simplifies Eq. \ref{eq:Dtilde} to
\begin{equation}\label{eq:adif}
	\frac{d}{dt}\tilde{\DD}_{mn}(t) = \hspace{-5pt}\sum_{\substack{q\\E_q = E_n}}\tilde{\DD}_{mq}(t)M_{qn}(t) 
	- \hspace{-5pt}\sum_{\substack{p\\E_p = E_m}}M_{mp}(t)\tilde{\DD}_{pn}(t),
\end{equation}
which we study in two separate cases. First suppose that states $m$ and $n$ are non-degenerate. Under this condition, the sum restriction $E_q = E_n$ implies that $E_q \neq E_m$ because the contrary would contradict the assumption of non-degeneracy ($E_m \neq E_n$). The same applies for the $p$ sum so that only terms depending on $\DD_{ij}(t)$ with $E_i(s) \neq E_j$ appear on the right hand side. If we choose matrices with $D_{ij}(0) = 0$ for all states $i,j$ with $E_i \neq E_j$, these terms remain $D_{ij}(t) = 0$ at all times because these elements only intermix with each other.

What remains is to evaluate Eq. \ref{eq:adif} when $m$ and $n$ are degenerate. In this case, $p$ and $q$ both run over all states degenerate with $m$ and $n$ so that the right-hand side reduces to a single sum. We see from Eq. \ref{eq:dynphase} that the degeneracy condition implies that $\tilde{\boldsymbol{\DD}}(t) = \boldsymbol{\DD}(t)$. Taking the basis to be ordered by increasing $E_i(t)$, we can define submatrices $\bs{\DD}_\mu(t)$ for each degenerate subspace $\mu$. Noting that the terms in Eq. \ref{eq:adif} are in the form of matrix multiplications, we find that the equation of motion separates into the commutation relations
\begin{equation}\label{eq:Ddot}
	\dot{\boldsymbol{\DD}}_\mu(t) = -[\mathbf{M}_\mu(t),\boldsymbol{\DD}_\mu(t)]
\end{equation}
for each subspace $\mu$. This expression details the accumulation of geometric phase within a given degenerate subspace as time progresses. The final GDM is simply the block diagonal direct sum
\begin{equation}\label{eq:Dsum}
	\boldsymbol{\DD}(t) = \text{diag}[\boldsymbol{\DD}_1(t),\boldsymbol{\DD}_2(t),\dots,\boldsymbol{\DD}_\mu (t),\dots],
\end{equation}
valid for all times including $t>T$. As a final remark, note that Eq. \ref{eq:Ddot} describes a unitary transformation because the matrix $\mathbf{M}(t)$ is skew-Hermitian. We may also choose to define a Hermitian matrix $\tilde{\mathbf{M}}(t) = -i\mathbf{M}(t)$ so that the expression more closely resembles Eq. \ref{eq:Lvn}.

\subsubsection{Practical considerations of the degenerate adiabatic theorem}

The application of Eqs. \ref{eq:Ddot} and \ref{eq:Dsum} requires knowledge of the matrix $\mathbf{M}(t)$. Using the relation $(d/dt)=\dot{\lambda}(d/d\lambda)$, we can calculate the matrix elements from Eq. \ref{eq:Mel} in the position representation
\begin{equation}\label{eq:connection}
	M_{ij}(t) = -i \dot{\lambda}(t)\int \psi_i^*(\xx|\lambda)\frac{\partial}{\partial\lambda}\psi_j(\xx|\lambda)d\xx.
\end{equation}
However, our only prescription for the instantaneous eigenstates was that they diagonalized the Hamiltonian at time $t$. We have not fixed the connection between eigenstates $\psi_i(\xx|\lambda)$ and $\psi_i(\xx|\lambda + \Delta \lambda)$. In fact, an eigenstate solver may output a different arbitrary rotation for each value of $\lambda$ so that $\psi_j(\xx|\lambda)$ is discontinuous in $\lambda$ and the $T\rightarrow \infty $ limit of Eqs. \ref{eq:Ip} is no longer valid. Thus, we must require that $\psi_j(\xx|\lambda)$ is smooth as a function of $\lambda$. In Section \ref{sec:Solve} we will perturb the Hamiltonian with an asymmetric potential to lift the degeneracy so that the connection between time-adjacent eigenstates is obvious.

The final issue pertains to the $\lambda$-dependent spectrum of geminal eigenstates. As eigenvalues approach each other with increasing $\lambda$, they may cross or anti-cross depending on symmetry. Anti-crossings are handled perfectly fine by Eqs. \ref{eq:Ddot} and \ref{eq:Dsum}, but the derivation of the adiabatic theorem breaks down at a crossing point. In a sense, the point acts as a pole where the degenerate submatrices intermix by instantaneously picking up geometric phase within the expanded degenerate subspace at the intersection.

We determine the behavior around these accidental degeneracies by a physical argument. First note that the total energy as computed by the many-body wave function must be differentiable, which we see upon differentiating the energy expectation value by the Hellman-Feynman theorem:
\begin{equation}\label{eq:finite}
	\frac{d\EE}{dt} = \dot{\lambda}(t)\sum_{i,j>i}\int\frac{|\Psi(\overline{\xx}|t)|^2}{|\mathbf{r}_i - \mathbf{r}_j|}d\overline{\xx}.
\end{equation}
It follows that $\EE(t)$ is differentiable because the right hand side of Eq. \ref{eq:finite} is finite at all times. The GDM that represents the wave function computes the energy as 
\begin{equation}\label{eq:Etime}
	\EE(t) = \sum_{\mu}\Tr[\boldsymbol{\mathcal{D}}_{\mu}(t)]E_\mu(\lambda(t)),
\end{equation}
where $\Tr[\boldsymbol{\mathcal{D}}_{\mu}(t)]$ is the population in a given degenerate subspace with energy $E_\mu(\lambda(t))$. An abrupt re-distribution of population at a crossing point $t_0$ would yield an $\EE(t)$ that is not differentiable at $t_0$. Since we know by Eq. \ref{eq:finite} that $\EE(t)$ must be differentiable, we conclude that no instantaneous population interchange can occur at a level crossing.

\section{Computing many-body eigenstates using the adiabatic theorem}\label{sec:Solve}
We are now equipped to solve the many-electron Schr\"odinger equation
\begin{equation}\label{eq:Hfin}
	\left(\sum_{ij}H(\mathbf{x}_i,\mathbf{x}_j)\right)\Psi_n(\overline{\xx}) = \mathcal{E}_n \Psi_n(\overline{\xx})
\end{equation}
for the matrices $\mathbf{D}_n$ that represent wave functions $\Psi_n(\overline{\xx})$. The Hamiltonian of interest has terms 
\begin{align}\label{eq:Hijf}
	H(\mathbf{x}_i,\mathbf{x}_j) =  H_1(\mathbf{x}_i,\mathbf{x}_j) + \frac{1}{|\mathbf{r}_i - \mathbf{r}_j|},
\end{align}
containing the one-body contribution
\begin{equation}
	H_1(\mathbf{x}_i,\mathbf{x}_j) = \frac{1}{N-1}\left[-\frac{\nabla_i^2}{2} - \frac{\nabla_j^2}{2} + v(\mathbf{x}_i) + v(\mathbf{x}_j)\right]
\end{equation}
along with the Coulomb potential $1/|\mathbf{r}_i - \mathbf{r}_j|$. For reasons that will be explained shortly, we treat Eqs. \ref{eq:Hfin} and \ref{eq:Hijf} as a special case of the two-parameter Hamiltonian
\begin{align}\label{eq:Vpert}
	H(\epsilon,\lambda) &= \sum_{i,j>i}\Bigg[ H_1(\mathbf{x}_i,\mathbf{x}_j) \nonumber \\
	&+ \epsilon \left(\frac{v_p(\mathbf{x}_i) + v_p(\mathbf{x}_j)}{N-1}\right) + \lambda \left(\frac{1}{|\mathbf{r}_i - \mathbf{r}_j|}\right)\Bigg]
\end{align}
that reduces to the original operator when $\epsilon = 0$ and $\lambda = 1$. The one-body perturbing potential $v_p(\mathbf{x})$ is chosen to have symmetry group containing only the identity so that the spectrum of the auxiliary many-body system
\begin{equation}\label{eq:H2p}
	H(\overline{\xx}|\epsilon,\lambda)\Psi_n(\overline{\xx}|\epsilon,\lambda) = \EE_n(\epsilon,\lambda)\Psi_n(\overline{\xx}|\epsilon,\lambda)
\end{equation}
is nondegenerate when $\epsilon \neq 0$. 


%

We represent a given eigenstate $\Psi_n(\overline{\xx}|\epsilon,\lambda)$ of Eq. \ref{eq:H2p} by its GDM $\boldsymbol{\DD}(\epsilon,\lambda)$, dropping the subscript $n$. The electronic energy is then calculated by
\begin{equation}\label{eq:Esumup}
	\EE_n(\epsilon,\lambda) = \sum_{i}\DD_{ii}(\epsilon,\lambda)E_i(\epsilon,\lambda),
\end{equation}
where each $E_i(\epsilon,\lambda)$ is an eigenvalue of the effective two-particle Hamiltonian $H(\xx|\epsilon,\lambda)$ satisfying the Schr\"odinger equation
\begin{equation}\label{eq:g2e}
	H(\xx|\epsilon,\lambda)\psi_i(\xx|\epsilon,\lambda) = E_i(\epsilon,\lambda)\psi_i(\xx|\epsilon,\lambda)
\end{equation}
for the geminal eigenstate $\psi_i(\xx|\epsilon,\lambda)$.

As mentioned, the analysis takes parameters $\epsilon$ and $\lambda$ to be functions of time. The dependence is chosen in such a way that the Hamiltonian slowly changes from non-interacting to fully interacting over some time interval. Beginning at $t=0$ with $\epsilon(0) = \lambda(0) = 0$, we slowly ramp the symmetry-breaking parameter $\epsilon(t)$ over time $T_1$ so that $\epsilon(T_1) = 1$ and $\lambda(T_1) = 0$. The nondegenerate eigenstates at $T_1$ remain Slater determinants due to the absence of electron-electron interactions. 

\begin{figure}[h]
	\includegraphics{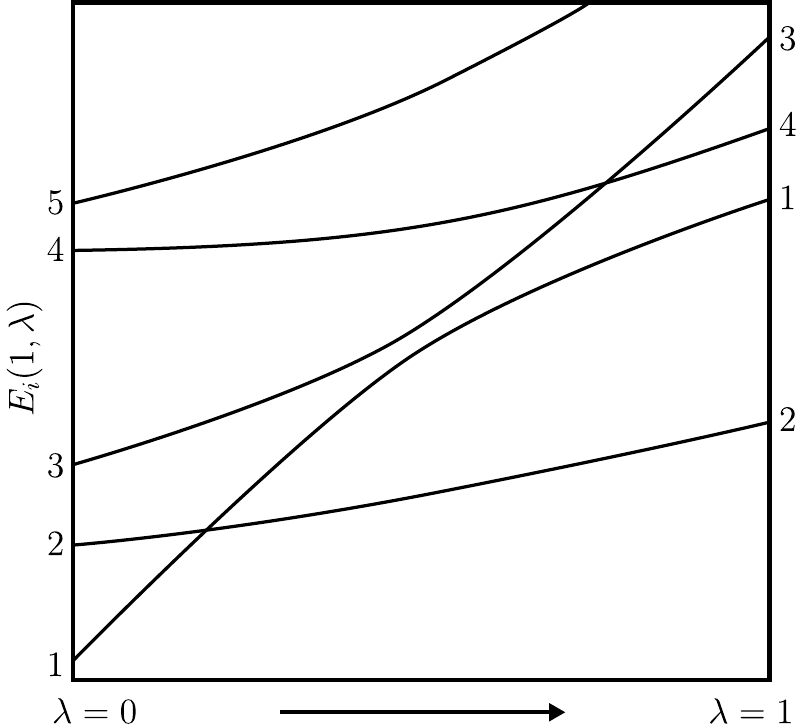}
	\caption{Example illustration of the eigenvalues of $H(\xx|1,\lambda)$ versus $\lambda$. $\lambda = 0$ and $\lambda = 1$ eigenstate indices are marked by numbers on the vertical axes. The curves do not represent the diagonalization of a real Hamiltonian.}
	\label{fig:adi}
\end{figure}

The crucial next step is to smoothly switch on the Coulomb interaction by increasing $\lambda(t)$ from $0$ at $T_1$ to $1$ at $T_2$ with $\epsilon = 1$. Assuming adiabatic evolution in which $T_2 - T_1 \rightarrow \infty$, we have by Eq. \ref{eq:Ddot} that the GDM follows the trivial relation $\dot{\boldsymbol{\DD}}(t) = 0$ because the two-electron spectrum is nondegenerate. In simple terms, all elements of the GDM stay fixed while the geminal eigenstates evolve to satisfy Eq. \ref{eq:g2e} for each $\lambda(t)$ with $\epsilon = 1$. Fig. \ref{fig:adi} is an illustration of how these parameter-dependent spectra may look. The curves do not represent the diagonalization of a real Hamiltonian, but they provide insight into how the technique works. 

For the simplest non-trivial example, consider a 3-electron state constructed from the geminal eigenstates of Fig. \ref{fig:adi}. Suppose that the GDM with geminals 1--3 occupied represents a valid Slater determinant at $\lambda(T_1) = 0$. As time progresses, the matrix elements remain unchanged as each eigenstate evolves to its $\lambda(T_2) = 1$ counterpart with $E_i(1,\lambda)$ varying smoothly with $\lambda$. The smoothness condition is imposed to define the behavior at level crossings; the total energy calculated by Eq. \ref{eq:Esumup} should be differentiable at all times by Eq. \ref{eq:finite}. As a consequence of the level crossing between eigenvalues 3 and 4, the $\lambda = 1$ (interacting) GDM with the 3 lowest energy eigenstates occupied is not $N$-representable if the configuration $\conf{1,2,4}$ is not a valid $\lambda = 0$ Slater determinant.

Generalizing to $N$ electrons, the principal task is to compute at least $N(N-1)/2$ geminal eigenstates on a grid of $\lambda$ points to generate curves like those in Fig. \ref{fig:adi}. This number of states is a lower bound due to constraints on the initial Slater determinant as well as the existence of level crossings. With sufficiently fine $\lambda$ sampling, it is possible to resolve all such crossings to connect the $\lambda = 0$ and $\lambda = 1$ geminals. This two-electron diagonalization can be performed with standard techniques like the configuration interaction, noting that the adiabatic theorem only serves to select which states are $N$-representable by virtue of their being the result of evolution under a time-dependent Hamiltonian. 

After determining valid symmetry-broken states by the above procedure, we must finally ramp $\epsilon$ to zero at time $T_3$ to discover solutions to the target Hamiltonian with $\epsilon = 0$ and $\lambda = 1$. During this process, the degenerate subspaces recombine as $\epsilon$ decreases. We can calculate the exact total energy at $T_3$ by 
\begin{equation}\label{eq:Efind}
	\EE = \sum_\mu \Tr[\boldsymbol{\DD}_\mu(0)]E_\mu(0,1),
\end{equation}
where $\Tr[\boldsymbol{\DD}_\mu(0)]$ is the time-invariant population of subspace $\mu$. Even though the Slater determinants were chosen at $T_1$, we may simply back-propagate to determine the population at the initial time. Repeating for various $t=T_1$ Slater determinants, the ground state of the interacting system is that which yields the minimum energy by Eq. \ref{eq:Efind}.

For observables other than the energy, the best we can do is calculate the $\epsilon$-dependent expectation value by 
\begin{equation}
	\braket{A}(\epsilon) = \Tr[\boldsymbol{\DD}(T_1) \mathbf{A}(\epsilon)].
\end{equation}
The matrix $\mathbf{A}$ can be determined to arbitrary precision by computing it from the geminal eigenstates with sufficiently small $\epsilon$. Conceptually, this limiting procedure of switching on and off the perturbation $\epsilon$ can be considered to account for the geometric phase that would be picked up in a degenerate adiabatic evolution. 

\subsection{Calculating atomic energy eigenstates}
This section details a simple but powerful example of the calculation of many-body eigenstates using the GDM. We will find that the many-body Schr\"odinger equation for an arbitrary atom or ion is analyzable strictly through the solution of an appropriately-scaled Helium atom problem.

Begin with the Hamiltonian $H_{Z,N}$ for a central potential with nuclear charge $Z$ and $N$ electrons. The fully-interacting system obeys the effective two-particle Hamiltonian of Eq. \ref{eq:Ht} with $\lambda = 1$: 
\begin{align}\label{eq:Hzn}
	H_{Z,N}(\mathbf{r}_1,\mathbf{r}_2|1) &= \frac{1}{N-1}\Bigg[-\frac{\nabla_1^2}{2} - \frac{\nabla_2^2}{2} \nonumber \\
	&- Z\left(\frac{1}{|\mathbf{r}_1|} + \frac{1}{|\mathbf{r}_2|}\right) + \frac{N-1}{|\mathbf{r}_1 - \mathbf{r}_2|}\Bigg].
\end{align}
It will be convenient to reduce the bracketed expression into a form that resembles the Helium atom Hamiltonian
\begin{equation}\label{eq:H22}
	H_{2,2}(\mathbf{r}_1,\mathbf{r}_2|\lambda) = -\frac{\nabla_1^2}{2} - \frac{\nabla_1^2}{2} - \frac{2}{|\mathbf{r}_1|} - \frac{2}{|\mathbf{r}_2|} + \frac{\lambda}{|\mathbf{r}_1 - \mathbf{r}_2|}.
\end{equation}
for some choice of $\lambda$. We find this simplification by transforming the coordinates into a yet-undetermined natural scale
\begin{equation}
	\overline{\mathbf{r}} = a \mathbf{r}.
\end{equation}
This transformation changes Eq. \ref{eq:Hzn} to
\begin{align}\label{eq:Hznt}
	H_{Z,N}(\overline{\mathbf{r}}_1,\overline{\mathbf{r}}_2|1) &= \frac{a^2}{N-1}\Bigg[-\frac{\overline{\nabla}_1^2}{2} - \frac{\overline{\nabla}_2^2}{2} \nonumber \\
	&-\frac{Z}{a}\left(\frac{1}{|\overline{\mathbf{r}}_1|} + \frac{1}{|\overline{\mathbf{r}}_2|}\right)  + \frac{N-1}{a}\frac{1}{|\overline{\mathbf{r}}_1 - \overline{\mathbf{r}}_2|}\Bigg].
\end{align}
Eq. \ref{eq:Hznt} reduces to a scaled copy of the Helium Hamiltonian in Eq. \ref{eq:H22} if we choose
\begin{equation}
	a = \frac{Z}{2},
\end{equation}
from which it follows that $\lambda = 2(N-1)/Z$. In the end we relate the $Z,N$ and $2,2$ (Helium) Hamiltonians by
\begin{equation}\label{eq:Hzn22}
	H_{Z,N}(\overline{\mathbf{r}}_1,\overline{\mathbf{r}}_2|1) = \left(\frac{Z}{2}\right)^2\frac{1}{N-1} H_{2,2}\Bigg(\overline{\mathbf{r}}_1,\overline{\mathbf{r}}_2\Bigg|\frac{2(N-1)}{Z}\Bigg).
\end{equation}
Eq. \ref{eq:Hzn22} allows us to diagonalize any atomic Hamiltonian by solving for eigenstates of the Helium atom with the appropriate values of $\lambda$. Accounting for coordinate scaling, the geminal basis functions are
\begin{equation}\label{eq:psizn}
	\psi_{Z,N}(\mathbf{r}_1,\mathbf{r}_2|1) = \psi_{2,2}\left(\frac{Z\mathbf{r}_1}{2},\frac{Z\mathbf{r}_2}{2}\right.\left|\frac{2(Z-1)}{Z} \right).
\end{equation}
Eqs. \ref{eq:Hzn22} and \ref{eq:psizn} can be used in conjunction with the two-parameter solution method outlined in Section \ref{sec:Solve}. Finally, we observe that for any neutral atom with $Z=N$ it suffices to calculate $\sim N(N-1)/2$ eigenstates of $H_{2,2}(\overline{\mathbf{r}}_1,\overline{\mathbf{r}}_1|\lambda)$ for $\lambda$ between $0$ and $2$. 

\section{Discussion}
We introduced the geminal density matrix as the basic variable to describe a many-body quantum state. The GDM was found to faithfully represent the $N$-electron wave function because it reproduces the expectation value of an arbitrary observable. The GDM is remarkably simple because it derives directly from the exchange symmetries of the $N$-electron wave function and a given operator $A$ without postulating the existence of statistical ensembles. Furthermore, it is analyzed strictly by matrix operations without needing higher order tensors.

The key behavior of the GDM is that it evolves by unitary matrix transformation, which we derived by assuming that the two-body contributions of the Hamiltonian and operator $A$ depended only on position. This likely imposes no limitations on the applicability of the GDM because it is unclear whether two-body interactions with derivatives exist. We can certainly solve for the eigenstates of a system with Coulomb repulsion along with more exotic interactions like the attractive potential mediated by lattice vibrations in the theory of superconductivity. In the case that two-body derivative terms exist, we must restrict the valid operators for which the expectation value is computable to include only one-body operators and the Hamiltonian itself.

The simple time evolution law allowed us to recover the classical intuition of electron-nuclei thermalization from an exact quantum-mechanical treatment of the electron gas. We continued to derive a degenerate adiabatic theorem which we exploited to calculate the stationary states of an arbitrary many-body Hamiltonian. We found that the problem of directly minimizing the energy functional for an $N$-electron wave function reduces to the calculation of around $N(N-1)/2$ eigenstates of an effective two-electron Hamiltonian on a grid of electron-electron interaction scaling strengths.

We finally displayed the power of this diagonalization method by applying it to atomic Hamiltonians, which reduced to an analysis of an effective Helium atom. So long as we know $\sim Z(Z-1)/2$ eigenstates of the Helium atom with coulomb interaction scaled by $\lambda$ from $0$ to $2$, we are able to compute the exact eigenstates of any atom. While the solution to the two-body problem is not trivial, it provides an incredible speedup for the solution of the many-body problem.

\acknowledgments
I thank Jens Biegert for the fruitful conversations that inspired this work along with his guidance and support throughout its completion. I also thank Eric Van Stryland and David Hagan for critical readings of the manuscript and acknowledge funding from the US Fulbright Student Program and the Air Force Office of Scientific Research grant FA9550-20-1-0322.

\appendix
\section{Matrix Constraints}\label{sec:Properties}
This appendix derives constraints on the form of $\mathbf{D}$ that follow directly from the definitions (Eq. \ref{eq:rho1})
\begin{equation}\label{eq:rhoapp}
	\rho(\xx,\xx') = \int \Psi^*(\xx,\yy)\Psi(\xx',\yy) d\yy
\end{equation}
and (Eq. \ref{eq:rhoexp})
\begin{equation}\label{eq:expapp}
	\rho(\xx,\xx') = \sum_{mn}D_{mn}\psi^*_n(\xx)\psi_m(\xx').
\end{equation}
These restrictions will serve as necessary $N$-representability conditions for the GDM to represent a valid $N$-electron wave function. 

The first rule arises from Eq. \ref{eq:rhoapp} which has the property $\rho(\xx',\xx) = \rho^*(\xx,\xx')$. Applying this transformation directly to the expansion in Eq. \ref{eq:expapp} yields
\begin{align} \label{eq:rho2star}
	\sum_{mn} D_{mn} \psi_n^*(\xx')\psi_m(\xx)\nonumber = \sum_{mn} D_{mn}^* \psi_n(\xx)\psi_m^*(\xx').
\end{align}
After swapping the sum index labels on the right hand side, we find the symmetry to be satisfied when $D_{nm}^* = D_{mn}$. This relationship implies the matrix identity
\begin{equation}\label{eq:Herm}
	\mathbf{D}^\dagger = \mathbf{D}.
\end{equation}

We uncover another $N$-representability requirement by fixing $\xx' = \xx$ in Eq. \ref{eq:rhoapp} and integrating both sides over the remaining free coordinates $\xx$. The integral on the right-hand side reduces to unity by the normalization condition of the wave function. Choosing a convenient representation for the left hand side gives
\begin{equation}\label{eq:tr2}
	\int\delta(\xx - \xx')\rho(\xx,\xx')d\xx d\xx' = \begin{pmatrix}
		N \\ 2
	\end{pmatrix}.
\end{equation}
Once again expanding $\rho(\xx,\xx')$ by Eq. \ref{eq:expapp}, we find using the orthonormality of the  geminal basis that
\begin{align}
	\Tr[\mathbf{D}] = \begin{pmatrix}
		N \\ 2
	\end{pmatrix}.
\end{align}

The requirement for antisymmetry under the exchange $\mathbf{x}_1 \leftrightarrow \mathbf{x}_2$ (or $\mathbf{x}_1' \leftrightarrow \mathbf{x}_2'$) follows from that of the many-body wave function, so that $\rho(\mathbf{x}_2,\mathbf{x}_1,\xx') = -\rho(\xx,\xx')$. Swapping these coordinates in the two-body expansion of Eq. \ref{eq:expapp} gives
\begin{align}
	\rho(\mathbf{x}_2,\mathbf{x}_1,\xx') &= \sum_{mn} D_{mn} \psi_n^*(\mathbf{x}_2,\mathbf{x}_1)\psi_m(\xx')\nonumber \\
	&= -\rho(\xx,\xx'),
\end{align}
indicating that the property is inherited from the anti-symmetry of the basis functions and does not further restrict $\mathbf{D}$. 

Unfortunately, we have now found all the $N$-representability conditions that follow directly from Eqs. \ref{eq:rhoapp} and \ref{eq:expapp}.  To further understand the $N$-representability problem we must choose a basis and derive expressions for the matrix elements $D_{mn}$. We compute these matrix elements by pre-multiplying Eq.  \ref{eq:expapp} by $\psi_n(\xx)\psi_m^*(\xx')$ and integrating over $d\xx$ and $d\xx'$ to find
\begin{equation}\label{eq:Dmn1}
	D_{mn} = \int \psi_n(\xx)\psi_m^*(\xx') \rho(\xx,\xx') d\xx d\xx'.
\end{equation}
Continuing to substitute Eq. \ref{eq:rhoapp} into Eq. \ref{eq:Dmn1} yields the simple equation
\begin{equation}\label{eq:Dmn2}
	D_{mn} = \int \Theta_n^*(\yy)\Theta_m(\yy) d\yy,
\end{equation}
with overlap functions $\Theta_m(\yy)$ defined to be
\begin{equation}\label{eq:Theta}
	\Theta_m(\yy) = \int \psi_m^*(\xx) \Psi(\xx,\yy) d\xx.
\end{equation}

We choose the geminal basis functions $\psi_i(\xx)$ to be those formed by the anti-symmetrized product of two single particle spinors $\phi_i(\mathbf{x})$. Grouping the two index labels into the symbol $\mathbf{n} = \conf{n_1,n_2}$, the basis functions take the form
\begin{equation}\label{eq:basis}
	\psi_{\mathbf{n}}(\xx) = \frac{1}{\sqrt{2}}\left[\phi_{n_1}(\mathbf{x}_1)\phi_{n_2}(\mathbf{x}_2) - \phi_{n_2}(\mathbf{x}_1)\phi_{n_1}(\mathbf{x}_2)\right].
\end{equation}
Because our basis functions are labeled by two integers, our $2$-RDM expansion will temporarily take the form of a rank four tensor with components $D_{\mathbf{mn}}$. It will eventually be necessary to map each $\mathbf{n}$ to a single integer index to flatten this tensor into a matrix (see the Table \ref{tab:map} and the surrounding discussion).

The most general wave function $\Psi(\xx,\yy)$ that will appear in Eq. \ref{eq:Theta} is a possibly infinite linear superposition of $N$-body Slater determinants
\begin{equation}\label{eq:confexp}
	\Psi(\xx,\yy) = \sum_{\conf{\alpha}}C_{\conf{\alpha}}\Psi_{\conf{\alpha}}(\xx,\yy).
\end{equation}
As in the main body of this text, we defined a configuration $\conf{\alpha}$ to be an ordered list of single-particle spinors present in a given product of states. The Slater determinants are formed by the antisymmetrization operator
\begin{equation}\label{eq:symop1}
	\Psi(\mathbf{x}_1,\dots,\mathbf{x}_N) = \hat{S}_- \prod_{i=1}^N \phi_{\alpha_i}(\mathbf{x}_i)
\end{equation}
equivalent to the determinant expression
\begin{equation}\label{eq:slater}
	\Psi_{\conf{\alpha}}(\xx,\yy) =  \frac{1}{\sqrt{N!}}\left|\begin{matrix}
		\phi_{\alpha_1}(\mathbf{x}_1) & \phi_{\alpha_2}(\mathbf{x}_1) & \dots & \phi_{\alpha_N}(\mathbf{x}_1) \\
		\phi_{\alpha_1}(\mathbf{x}_2) & \phi_{\alpha_2}(\mathbf{x}_2) &  \dots & \phi_{\alpha_N}(\mathbf{x}_2) \\
		\vdots & \vdots & \ddots & \vdots \\
		\phi_{\alpha_1}(\mathbf{x}_N) & \phi_{\alpha_2}(\mathbf{x}_N) & \dots & \phi_{\alpha_N}(\mathbf{x}_N)
	\end{matrix}\right|.
\end{equation}

We now expand each $\Psi_{\conf{\alpha}}(\xx,\yy)$ in Eq. \ref{eq:confexp} along minors of the top two rows of Eq. \ref{eq:slater} to isolate the $\mathbf{x}_1$ and $\mathbf{x}_2$ dependence. The result is
\begin{widetext}
	\begin{align}\label{eq:fullexpansion}
		\Psi(\xx,\yy) = \frac{1}{\sqrt{N(N-1)}}\sum_{\conf{\alpha}}C_{\conf{\alpha}}\sum_{i,j>i}(-1)^{i+j-1}\left[\phi_{\alpha_i}(\mathbf{x}_1)\phi_{\alpha_j}(\mathbf{x}_2) - \phi_{\alpha_j}(\mathbf{x}_1)\phi_{\alpha_i}(\mathbf{x}_2)\right]\Psi_{\conf{\alpha}_{ij}}(\yy),
	\end{align}
\end{widetext}
where the reduced configuration $\conf{\alpha}_{ij} = \conf{\alpha}\setminus \conf{\alpha_i,\alpha_j}$ is the set subtraction of $\alpha_i$ and $\alpha_j$ from the original list of states. The state $\Psi_{\conf{\alpha}_{ij}}(\yy)$ is the determinant of the matrix formed by removing rows $1$ and $2$ and columns $i$ and $j$ from Eq. \ref{eq:slater}. As the normalization of this $N-2$ electron state requires the prefactor $1/\sqrt{(N-2)!}$, we multiplied by its inverse which partially canceled with the $1/\sqrt{N!}$ prefactor.

Continuing to normalize the $\mathbf{x}_1$ and $\mathbf{x}_2$ dependence into a two-electron Slater determinant $\psi_{\alpha_i \alpha_j}(\xx)$, we finally have
\begin{align}\label{eq:Psi2final}
	\Psi(\xx,\yy) = \sqrt{\frac{2}{N(N-1)}}&\sum_{\conf{\alpha}}C_{\conf{\alpha}}\sum_{i,j>i}(-1)^{i+j-1} \nonumber \\ &\times\psi_{\alpha_i \alpha_j}(\xx)\Psi_{{\conf{\alpha}}_{ij}}(\yy).
\end{align}

Plugging Eq. \ref{eq:Psi2final} into Eq. \ref{eq:Theta} for the overlap $\Theta_{\mathbf{m}}(\yy)$, the integration over $d\xx$ reduces the two-electron wave functions to $\delta_{\alpha_i,m_1}\delta_{\alpha_j,m_2}$ by orthonormality. Thus, a given configuration that does not contain $\mathbf{m}=\conf{m_1,m_2}$ will not contribute to $D_{\mathbf{mn}}$. Consequently, we may reduce the $\conf{\alpha}$ (configuration) sum into one over $\conf{\alpha} \ni \conf{m_1,m_2}$. The remaining sum is reduced to the single term with $(i,j) = \conf{m_1,m_2}$ so that
\begin{align}\label{eq:Thetaf}
	\Theta_\mathbf{m}(\yy) = \sqrt{\frac{2}{N(N-1)}}\sum_{\conf{\alpha} \ni \mathbf{m}}&C_{\conf{\alpha}}\mathcal{S}_\alpha[\mathbf{m}] \Psi_{{\conf{\alpha}}_{\mathbf{m}}}(\yy),
\end{align}
where $\conf{\alpha}_{\mathbf{m}} = \conf{\alpha}_{m_1 m_2} = \conf{\alpha}\setminus \conf{m_1,m_2}$. The symbol $S_{\alpha}[\mathbf{m}]$ is the sign function
\begin{equation}
	S_{\alpha}[\mathbf{m}] = (-1)^{I_{\alpha}[m_1] + I_{\alpha}[m_2] - 1}
\end{equation}
with $I_\alpha[p]$ the index of basis function $p$ in configuration $\conf{\alpha}$. We absorb this sign into the expansion coefficient by defining $\mathcal{C}_{\conf{\alpha}} = C_{\conf{\alpha}}S_{\alpha}[\mathbf{m}]$.

Finally, using that $\Theta_n^*(\yy)$ is the complex conjugate of Eq. \ref{eq:Thetaf}, we compute $D_{\mathbf{mn}}$ by Eq. \ref{eq:Dmn2}:
\begin{align}
	D_{\mathbf{mn}} &= \sum_{\substack{\conf{\alpha} \ni \mathbf{m} \\ \conf{\beta} \ni \mathbf{n}}}\mathcal{C}^*_{\conf{\beta}}\mathcal{C}_{\conf{\alpha}}\int \Psi^*_{{\conf{\beta}}_{\mathbf{n}}}(\yy) \Psi_{{\conf{\alpha}}_{\mathbf{m}}}(\yy) d\yy.
\end{align}
The integral, being the inner product between orthonormal $N-2$ electron Slater determinants, equals one when $\conf{\alpha}_\mathbf{n} = \conf{\beta}_\mathbf{m}$ and zero otherwise. Therefore, 
\begin{align}\label{eq:Dndiag}
	D_{\mathbf{m}\mathbf{n}} &= \sum_{\substack{\conf{\alpha} \ni \mathbf{n} \\ \conf{\beta} \ni \mathbf{m}}}\mathcal{C}^*_{\conf{\beta} }\mathcal{C}_{\conf{\alpha}}\delta_{\conf{\alpha}_\mathbf{n}, \conf{\beta}_\mathbf{m}}.
\end{align}
The diagonal matrix elements are then found from Eq. \ref{eq:Dndiag} to take the simple form
\begin{align}\label{eq:Dmnfinal}
	D_{\mathbf{n}\mathbf{n}} = \sum_{\conf{\alpha}\ni \mathbf{n}}\left|C_{\conf{\alpha}}\right|^2.
\end{align}
Because Eq. \ref{eq:Dmnfinal} is a sum over the magnitude squared of all expansion coefficients of configurations containing $\mathbf{n}$, the overall normalization condition $\sum_{\conf{\alpha}}|C_{\conf{\alpha}}|^2 = 1$ implies that 
\begin{equation} \label{eq:Dnnb}
	0 \leq D_{\mathbf{nn}} \leq 1.
\end{equation}

The maximum diagonal value, $D_{\mathbf{nn}} = 1$, occurs when every configuration $\conf{\alpha}$ contains $\mathbf{n}$. In this case we encounter the additional rule that
\begin{equation}\label{eq:offdiag}
	D_{\mathbf{nn}} = 1 \implies \forall \mathbf{m} \neq \mathbf{n}, D_{\mathbf{mn}} = D_{\mathbf{nm}} = 0, 
\end{equation}
meaning that $1$ on the diagonal in position $\mathbf{n}$ forces all other elements in the column and row $\mathbf{n}$ to zero. 

The proof of Eq. \ref{eq:offdiag} proceeds as follows. Per Eq. \ref{eq:Dmnfinal}, any non-zero term must simultaneously satisfy the conditions $\conf{\alpha}\ni \conf{m_1,m_2}$, $\conf{\beta}\ni \conf{n_1,n_2}$ and $\conf{\alpha}\setminus \conf{m_1,m_2} = \conf{\beta} \setminus \conf{n_1,n_2}$. By this equality, $\conf{\alpha} \setminus \conf{m_1,m_2}$ does not contain $\conf{n_1,n_2}$. Since $\conf{\alpha}$ is formed by the set addition of some $\conf{m_1,m_2}\neq \conf{n_1,n_2}$, we have that $\conf{\alpha} \not \ni \conf{n_1,n_2}$. Supposing now that $D_{\mathbf{mn}} \neq 0$ implies existence of some $\conf{\alpha}$ in the state expansion that does not contain $\conf{n_1,n_2}$. This contradicts the requirement that must be met for $D_{\mathbf{nn}} = 1$ so we conclude that the existence of $1$ on a diagonal implies all other elements in that row and column are $0$.

By the Hermiticity (Eq. \ref{eq:Herm}) of $\mathbf{D}$, it can always be transformed into diagonal form by a unitary basis transformation (see Appendix \ref{sec:Basis}). The resulting diagonal matrix obeying Eq. \ref{eq:Dnnb} has the property
\begin{equation}\label{eq:trsquared}
	0\leq\Tr[\mathbf{D}^2] \leq \nc,
\end{equation}
which follows trivially from the fact that $a^2 \leq a$ for a number $a \leq 1$. $\Tr[\mathbf{D}^2]$ is a basis-indepedent quantity as it is invariant under unitary transformation by the cyclic property of the trace.

We finally summarize the necessary $N$-representability constraints on the matrix $\mathbf{D}$:
\begin{subequations}
	\label{eq:rules}
	\begin{align}
		\mathbf{D} &= \mathbf{D}^\dagger \label{eq:hermitian}\\
		0\leq &D_{nn} \leq 1 \label{eq:occupation}\\
		\Tr[\mathbf{D}] &= \begin{pmatrix}
			N \\ 2
		\end{pmatrix}  \label{eq:trace}\\
		0\leq \Tr[\mathbf{D}^2] &\leq \begin{pmatrix}
			N \\ 2
		\end{pmatrix}. \label{eq:tracesquared}
	\end{align}
\end{subequations}

\section{Change of basis}\label{sec:Basis}
The formula for a change of basis is identical to the transformation for any density matrix, but we re-derive it here for completeness. We begin with the two-body density matrix
\begin{align}
	\mathbf{D} = \sum_{mn}D_{mn}\psi^*_{n}(\xx)\psi_m(\xx)
\end{align}
and introduce a new orthonormal basis with wave functions $\phi_i(\xx)$. Expanding the initial states $\psi_n(\xx)$ in terms of the new, we find
\begin{align}\label{eq:changebasis}
	\mathbf{D} &= \sum_{mn}D_{mn}\left(\sum_{j}U^*_{jn}\phi^*_{j}(\xx)\right)\left(\sum_i U_{im} \phi_i(\xx')\right)\nonumber \\
	&= \sum_{ij}\left(\sum_{mn} U_{im}D_{mn}U^*_{jn}\right)\phi^*_j(\xx)\phi_i(\xx').
\end{align}
The parenthetical term gives the expression for $D_{ij}$, which we take to be the coefficients of matrix $\mathbf{D}'$. We finally find the matrix form for the change of basis
\begin{equation}\label{eq:basischange}
	\mathbf{D}' = \mathbf{U}\mathbf{D}\mathbf{U}^\dagger,
\end{equation}
where $\mathbf{U}$ is the unitary matrix of coefficients $U_{ij}$.

\section{Alternate derivation of the GDM evolution equation}\label{sec:Alternate}

We found the matrix Liouville-von Neumann equation (Eq. \ref{eq:Lvn}) by deriving its most general operator equivalent then specializing to the case of a time-independent basis. We can perform an alternate derivation by starting with the time-independent expansion of the $2$-RDM
\begin{equation}\label{eq:rhont2}
	\rho(\xx,\xx'|t) = \sum_{mn}D_{mn}(t)\psi_n^*(\xx)\psi_m(\xx').
\end{equation}
We further assume that our dummy operator $A(\xx)$ is time independent. Noting that $(d/dt)\braket{A}(t) = -i K(t)$ for a time-independent $A(\xx)$, we restart from Eq. \ref{eq:Kwf},
\begin{equation}\label{eq:ddtapp}
	\frac{d}{dt}\braket{A}(t) = -i\int \psi_n^*(\overline{\xx}|t)[H(\xx|t),A(\xx)]\psi_m(\overline{\xx}|t)d\overline{\xx}.
\end{equation}
Expressing Eq. \ref{eq:ddtapp} in terms of the $2$-RDM as
\begin{align}\label{eq:KK2}
	\frac{d}{dt}\braket{A}(t) &= \int  d\xx d\xx' \delta(\xx-\xx') \nonumber \\
	&\times\left[H'(\xx|t),A(\xx)\right]\rho(\xx,\xx'|t),
\end{align}
we can apply the expansion in Eq. \ref{eq:rhont2} to Eq. \ref{eq:KK2}. In terms of the abstract effective two-electron operators, the result is
\begin{equation}
	\frac{d}{dt}\braket{A}(t) = \sum_{mn}D_{mn}[\braket{m|\hat{H}'(t)\hat{A}|n} -\braket{m|\hat{A}\hat{H}'(t)|n}],
\end{equation}
into which we can insert the identity $1 = \sum_i \ket{i}\bra{i}$ to form the matrix equation
\begin{equation}\label{eq:trOa1}
	\frac{d}{dt}\braket{A}(t) = -i\Tr\left[\mathbf{D}(t)[\mathbf{H}'(t),\mathbf{A}]\right].
\end{equation}
For comparison, directly differentiating the trace relation $\braket{A}(t) = \Tr[\mathbf{D}(t)\mathbf{A}]$ gives
\begin{equation}\label{eq:KK3}
	\frac{d}{dt}\braket{A}(t) = \Tr[\dot{\mathbf{D}}(t)\mathbf{A}],
\end{equation}
whose result must agree with Eq. \ref{eq:trOa1}.

Using the cyclic property and linearity of the trace, we can rearrange Eq. \ref{eq:trOa1} as $\Tr[\mathbf{D}\mathbf{H}'\mathbf{A} - \mathbf{D}\mathbf{A}\mathbf{H}'] = \Tr[\mathbf{D}\mathbf{H}'\mathbf{A} - \mathbf{H}'\mathbf{D}\mathbf{A}]$ so that Eq. \ref{eq:trOa1} takes the form
\begin{equation}\label{eq:dtapfin}
	\frac{d}{dt}\braket{A}(t) = \Tr[-i[\mathbf{D}(t),\mathbf{H}'(t)]\mathbf{A}(t)].
\end{equation}
Comparing Eq. \ref{eq:dtapfin} to Eq. \ref{eq:KK3} yields the condition
\begin{equation}
	\dot{\mathbf{D}}(t) = -i[\mathbf{D}(t),\mathbf{H}'(t)]
\end{equation}
to ensure the GDM gives the same expectation values as the wave function.

\bibliography{ManyBody}

\end{document}